\documentclass[aps,prb,preprint,showpacs,superscriptaddress]{revtex4-1}
\usepackage{graphicx}
\usepackage{epsfig}
\usepackage{epstopdf}
\usepackage{float}
\usepackage{subfigure}
\usepackage{amsmath}
\usepackage[section]{placeins} 	
\usepackage{mathtools}
\usepackage{needspace}
\DeclareMathOperator{\cov}{cov}

\usepackage[section]{placeins}       
\usepackage{pgfplots}			
\usepackage{tikz}
\usepackage{wrapfig}
\usetikzlibrary{patterns}  		

\newcommand{\exciting}{{\usefont{T1}{lmtt}{b}{n}exciting}}

\begin{document}
\title{Thermoelastic properties of $\alpha$-iron from first-principles }
\author{Daniele Dragoni}
\affiliation{Theory and Simulations of Materials (THEOS), and National Center for Computational Design and Discovery of Novel Materials (MARVEL), \'Ecole Polytechnique F\'ed\'erale de Lausanne, 1015 Lausanne, Switzerland}
\author{Davide Ceresoli}
\affiliation{CNR Istituto di Scienze e Tecnologie Molecolari (CNR-ISTM), 20133, Milano, Italy}
\author{Nicola Marzari}
\affiliation{Theory and Simulations of Materials (THEOS), and National Center for Computational Design and Discovery of Novel Materials (MARVEL), \'Ecole Polytechnique F\'ed\'erale de Lausanne, 1015 Lausanne, Switzerland}

\date{\today}

\begin{abstract}
We calculate the thermomechanical properties of $\alpha$-iron, and in particular
its isothermal and adiabatic elastic constants, using first-principles 
total-energy and lattice-dynamics calculations, minimizing the
quasi-harmonic vibrational free energy under finite strain deformations.
Particular care is made in the fitting procedure for the static and temperature-dependent
contributions to the free energy, in discussing error propagation for the
two contributions separately, and in the verification and validation of 
pseudopotential and all-electron calculations.
We find that the zero-temperature mechanical
properties are sensitive to the details of the calculation strategy employed,
and common semi-local exchange-correlation functionals provide only fair
to good agreement with experimental elastic constants, while their
temperature dependence is in excellent agreement with experiments
in a wide range of temperature almost up to the Curie transition.
\end{abstract}
\pacs{}


\maketitle

\section{Introduction}
Elemental iron is a material of great scientific and economic interest: it's the major constituent
of steels, it determines the properties of the earth core, and its
complex phase diagram is driven by the subtle interplay between 
vibrational and magnetic contributions, making it particularly challenging to describe accurately
from first-principles. This is especially relevant as temperature increases, since magnetic
excitations become important and a dramatic change in the magnetic nature of the system takes place.
At ordinary pressures, iron turns from a BCC ferromagnet to a BCC paramagnet with
a second-order transition ($\alpha \to \beta$) at a Curie temperature of $\sim$ 1043~K.
This transition is then followed by
two structural transitions, a BCC$\to$FCC ($\beta \to \gamma$) transition at  1185~K
and a FCC$\to$BCC ($\gamma\to\delta$) transition at 1670~K, before finally melting at
$\sim$~1814~K.

First-principles simulations can be a key technique and a valid alternative to experiments 
in order to get accurate predictions of phase diagrams without the need of phenomenological
parameters, and become essential at conditions that are challenging to reproduce in
real life, like those inside the earth's core~\cite{Alfe}. 
For the case of pressure-temperature phase diagrams, zero-temperature
first-principles equations of state can be supplemented with finite-temperature vibrational entropies,
that can be derived directly from the knowledge of the phonon dispersions.
These latter can be calculated from finite differences, or more elegantly and less expensively with
density-functional perturbation theory (DFPT)~\cite{DFT,baroniRMP}. When coupled to the quasi-harmonic 
approximation~\cite{QHA} (QHA), these techniques allow to calculate thermal expansion
and vibrational properties at finite temperatures, often well above the Debye temperature~\cite{Pavone-Baro,Liu, Mounet,shobhana,Debernardi,shob-degir,sha}.
While there have been numerous first-principles calculations of elastic
properties of solids by either total energy, stress-strain~\cite{Elastic,Mehl} or density-functional
perturbation theory approaches~\cite{Baro-Giannoz,Vanderbilt-elastic,Vanderbilt-elastic2} a relatively limited number of them has been 
focused on to the thermo-mechanical properties of metals or minerals~\cite{renata1,Renata-thermoelastic,Wang,Oganov}.

In this paper, we calculate the adiabatic and isothermal finite-temperature elastic properties 
of ferromagnetic $\alpha$-iron at ambient pressure and 
in the temperature range of stability for ferromagnetic $\alpha$-iron using the
QHA and DFPT. 
We also carefully explore multidimensional fitting procedures for the static and vibrational 
contributions to the free energy, and analyze the quality of the fit and the source of errors of 
both contributions, providing a confidence interval of each elastic constant as a function of temperature.

The paper is organized as follows: in Sec.~\ref{sec:framework} we introduce the finite strain
framework used to calculate the elastic constants, and in
Sec.~\ref{sec:compdetails} we give the computational details of our first-principles density-functional theory (DFT) and DFTP calculations.
We present our results and their comparison to experiments in Sec.~\ref{sec:results}.
Finally, Section~\ref{sec:conclusions} is devoted to summary and conclusions.

\section{Finite-strain method}
\label{sec:framework}
In the limit of small deformations, the energy of a crystal at an arbitrary configuration can be Taylor-expanded 
in terms of a symmetric
matrix $\boldsymbol{\varepsilon}$ describing a uniform linear deformation $\boldsymbol{A}$ such that
\begin{align}
  \boldsymbol{A}=\boldsymbol{1}+\boldsymbol{\varepsilon}
  \label{eq:El-def}
\end{align} 
and any position vector $\bar r$ in the reference configuration is transformed into 
$(\boldsymbol{1+\varepsilon} )\cdot \bar r$. The isothermal elastic constants (SOECs) at zero
pressure are then defined as the second-order coefficients of this expansion according to~\cite{Wallace}
\begin{align}
  C_{ij}^T=\frac{1}{V_0} \frac{\partial^2 F}{\partial \varepsilon_i \partial \varepsilon_j} \bigg |_{V_0,T},
  \label{eq:El-def2}
\end{align}
where $F$ is the Helmholtz free-energy and $\varepsilon_i$, $\varepsilon_j, i,j=1\dots6$ are the components
of the strain tensor $\boldsymbol{\varepsilon}$ (we adopt here the Voigt notation). 
Also, note that the second derivatives are evaluated at the thermodynamic equilibrium 
configuration with volume $V_0$ and at constant temperature $T$.


For cubic crystals, like $\alpha$-iron, only three elastic constants are needed to completely determine
the stiffness tensor and, therefore, fully characterize the mechanical response of the system in the
linear elastic regime. As a consequence only three independent deformations are sufficient, and we choose here the hydrostatic, tetragonal and trigonal deformations, 
shown in Tab.~\ref{tab:deform}. Each deformation fully determines one of the cubic elastic
constants (or elastic moduli).

\begin{table}[h]
\centering
\begin{tabular}{ccccccc}
\hline\hline
   $\boldsymbol{\varepsilon}^{(i)}$ & $\varepsilon_1$ &$\varepsilon_2$ &$\varepsilon_3$ &$\varepsilon_4$ &$\varepsilon_5$ &$\varepsilon_6$ \\ 
\hline
  $\boldsymbol{\varepsilon}^{(1)}$   & $\varepsilon_a$ & $\varepsilon_a$ & $\varepsilon_a$ & 0 & 0 & 0 \\
  $\boldsymbol{\varepsilon}^{(2)}$   & 0         & 0       & $\varepsilon_c$ & 0 & 0 & 0  \\
  $\boldsymbol{\varepsilon}^{(3)*}$  & 0         & 0       & 0        & $\varepsilon_d/2$ & $\varepsilon_d/2$ & $\varepsilon_d/2$\\
\hline\hline
\end{tabular}
\caption{Deformations and corresponding strain vectors in the Voigt notation: $^{(1)}$hydrostatic,
$^{(2)}$tetragonal and $^{(3)}$trigonal deformations are governed by a single parameter. $^*$Note that the
trigonal deformation reported here is the first-order expansion of the full strain tensor
$\boldsymbol{\varepsilon}^{(3)}$ with non-zero off-diagonal terms. }
\label{tab:deform}
\end{table}

In order to compute finite-temperature properties and, therefore, to calculate the Helmholtz free energy $F$,
the vibrational contributions must be added to the static energy contributions. The QHA~\cite{QHA} provides an analytical expression for the 
vibrational contribution to the free energy:
\begin{align}
  F(\{ a_i\},T) &= \underbrace{\vphantom{\sum_{\boldsymbol{q},\lambda}}E_{stat}(\{ a_i\})}_{\text{Static}} +
   \underbrace{\frac{1}{2} \sum_{\boldsymbol{q},\lambda} \hbar \omega_{\boldsymbol{q},\lambda}(\{ a_i\}) }_{\text{ZPE}}+ \nonumber\\
  & + \underbrace{k_B T \sum_{\boldsymbol{q},\lambda} \ln \bigg(  1-e^{-\frac{\hbar \omega_{\boldsymbol{q},\lambda}(\{ a_i\})}{k_B T}} \bigg)}_{\text{Thermal}},
  \label{eq:QHA}
\end{align}
where the sum is performed over all the phonon modes $\lambda$ and all the phonon wave vectors $\textbf{q}$ spanning the Brillouin zone (BZ). Here, $k_B$ is the Boltzmann constant
and $\omega_{\boldsymbol{q},\lambda}$ the vibrational frequencies of the different phonon modes, where in the QHA their explicit dependence on the geometry of the system via the primitive lattice vectors $\{a_i\}$ is accounted for.
The vibrational part, coming directly from the analytic partition function of a
Bose-Einstein gas of harmonic oscillators, can be split into a
zero-point energy term plus a contribution which depends explicitly on the temperature $T$.
We neglect here the thermal electronic effects, because they are believed to be small 
compared to the quasi-harmonic 
vibrational contribution~\cite{sha,electro-phonon} in the range of stability of the $\alpha$ phase. Magnetic
effects are also not considered, except for the longitudinal relaxation of the total magnetic moment as a function of strain, but they are known to be important approaching the Curie point~\cite{neugebauer,Neugebauer-phonon,Fultz} and their influence on the elastic properties is briefly discussed in Sec.~\ref{sec:discussion} in the light of previous studies. 

Eq.~\ref{eq:El-def2} is used to calculate the isothermal elastic constants at finite temperature; however,
in order to compare results with experimental data obtained from ultrasonic measurements~\cite{Dever},
we also calculate the adiabatic elastic constants, using the following 
relations~\cite{Wallace}
\begin{align}
& C^{(S)}_{11}-C^{(T)}_{11} = C^{(S)}_{12}-C^{(T)}_{12} = \nonumber \\
& = B^{(S)}-B^{(T)} = \frac{T\, V\, \alpha^2 {B^{(T)}}^2}{C_V}, \label{eq:Isotoadia}  \\
&  C^{(S)}_{44}-C^{(T)}_{44} = 0, \nonumber 
\end{align}
where the heat capacity at constant volume $C_V$ and the volumetric thermal expansion coefficient $\alpha$
are both calculated from the QHA. The superscripts $(S)$ and $(T)$ define the adiabatic and isothermal 
elastic constants $C_{ij}$ and bulk modulus $B$ respectively.

\section{Computational details and pseudopotential selection}
\label{sec:compdetails}
We calculate the first-principles elastic constants using DFT, as implemented in the \texttt{PWSCF} and \texttt{PHONON} packages of the \textsc{Quantum-ESPRESSO}
distribution~\cite{QE} for the static and lattice dynamical calculations, respectively. The calculations
are spin-polarized and the magnetic moment is free to vary collinearly in order to minimize the total energy.
In all calculations the exchange-correlation effects have been treated within the generalized-gradient
approximation (GGA) with the PBE functional~\cite{PBE}. We use an ultrasoft
pseudopotential~\cite{Ultrasoft} (USPP) from \textsl{pslibrary.0.3.0}~\footnote{For iron, this is identical to 0.2.1},
which includes also $3s$ and $3p$
semicore states~\footnote{This pseudopotential is uniquely labeled as \texttt{Fe.pbe-spn-rrkjus$\_$psl.0.2.1.UPF} } 
(i.e. 16 valence electrons) along with a plane-wave basis with a wavefunction kinetic-energy cutoff of
90~Ry and a cutoff of 1080~Ry for the charge density.
We sampled the BZ with an offset $24\times24\times24$ Monkhorst-Pack $k$-mesh, with a
Marzari-Vanderbilt smearing~\cite{marzari_smearing} of 0.005~Ry. 

Phonon calculations were carried out for each deformation within DFPT~\cite{baroniRMP}: 
the dynamical matrix and its eigenvalues are calculated on a $4\times4\times4$ mesh of special points 
in the BZ and Fourier-interpolated on an extended $21\times21\times21$ grid for the integration of 
thermodynamic quantities.
We arrived at this computational setup (cutoff, smearing and BZ sampling) after a careful investigation of
the convergence of total energy and individual phonon frequencies for different deformations. Also, we verified
that individual total energies and phonon frequencies do change smoothly as a function of strain.

Since the choice of the pseudopotential is of primary importance for a clear comparison with computational and experimental data in the literature, 
it is worth to stress that the one used here has been chosen among different candidates from the \textsl{pslibrary}~\footnote{\protect\url{http://www.qe-forge.org/gf/project/pslibrary}} 
and \textsl{GBRV} library~\footnote{\protect\url{http://www.physics.rutgers.edu/gbrv/}}  to reproduce, as closely as possible, the all-electron FLAPW equilibrium lattice parameter, bulk modulus 
at 0~K and local magnetization obtained from independent groups.
Also, for the sake of completeness, we compare against results obtained using the \textsc{VASP} code and associated pseudopotentials~\cite{vasp}.
%
%
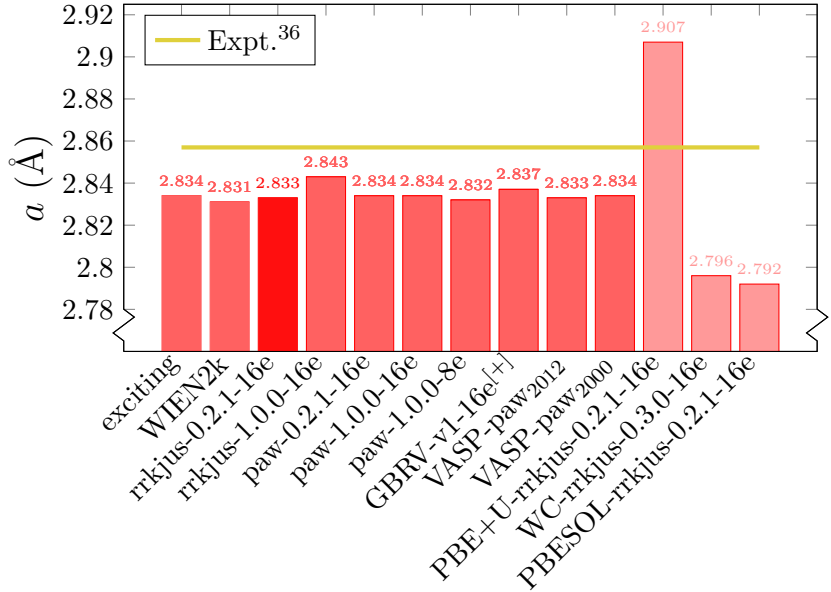
\begin {figure}[] 
\pgfplotsset{title=, width=10.8cm,height=6.2cm,}
\begin{center}
\begin{tabular}{rl}
\begin{tikzpicture} [baseline, trim axis right]]
\begin{axis}[
    ybar stacked,
    legend style={at={(0.5,1.15),font=\small},
    anchor=north west,legend columns=-1},
    ylabel={\large $a$  (\AA)},
    symbolic x coords={ exciting, WIEN2k, rrkjus-0.2.1-16e, rrkjus-1.0.0-16e, paw-0.2.1-16e, paw-1.0.0-16e, paw-1.0.0-8e, GBRV-v1-16e$^{[+]}$, VASP-paw$_{2012}$, VASP-paw$_{2000}$,PBE+U-rrkjus-0.2.1-16e,WC-rrkjus-0.3.0-16e, PBESOL-rrkjus-0.2.1-16e},
    xtick=data,
    x tick label style={rotate=45,anchor=east},
    ytick={ 2.78,2.80, 2.82,2.84,2.86,2.88,2.90,2.92 }, 
    ymin=2.76001, ymax=2.925, axis y discontinuity=crunch,
    bar width=5.2mm,
    every node near coord/.append style={font=\tiny,/pgf/number format/precision=3},
    nodes near coords 
    ]
\addplot+[ybar,red!62, draw=red] plot coordinates {  (exciting,0) (WIEN2k,0)  (rrkjus-0.2.1-16e,0) (PBE+U-rrkjus-0.2.1-16e,0) (rrkjus-1.0.0-16e,2.843) (paw-0.2.1-16e,2.834) (paw-1.0.0-16e,2.834) (paw-1.0.0-8e,2.832) (GBRV-v1-16e$^{[+]}$,2.837) (VASP-paw$_{2012}$,2.833) (VASP-paw$_{2000}$,2.834)  (WC-rrkjus-0.3.0-16e,0) (PBESOL-rrkjus-0.2.1-16e,0) };
\addplot+[ybar,red!62, postaction={pattern=north east lines}] plot coordinates { (exciting,2.834) (WIEN2k,2.831) (rrkjus-0.2.1-16e,0) (PBE+U-rrkjus-0.2.1-16e,0) (rrkjus-1.0.0-16e,0) (paw-0.2.1-16e,0) (paw-1.0.0-16e,0) (paw-1.0.0-8e,0) (GBRV-v1-16e$^{[+]}$,0) (VASP-paw$_{2012}$,0) (VASP-paw$_{2000}$,0) (WC-rrkjus-0.3.0-16e,0) (PBESOL-rrkjus-0.2.1-16e,0) };
\addplot+[ybar,red!94, postaction={pattern=crosshatch dots}] plot coordinates { (exciting,0)  (WIEN2k,0)  (rrkjus-0.2.1-16e,2.833) (PBE+U-rrkjus-0.2.1-16e,0) (rrkjus-1.0.0-16e,0) (paw-0.2.1-16e,0) (paw-1.0.0-16e,0) (paw-1.0.0-8e,0) (GBRV-v1-16e$^{[+]}$,0) (VASP-paw$_{2012}$,0) (VASP-paw$_{2000}$,0) (WC-rrkjus-0.3.0-16e,0) (PBESOL-rrkjus-0.2.1-16e,0) };
\addplot+[ybar,red!40, draw=red] plot coordinates {  (exciting,0) (WIEN2k,0)  (rrkjus-0.2.1-16e,0) (PBE+U-rrkjus-0.2.1-16e,2.907) (rrkjus-1.0.0-16e,0) (paw-0.2.1-16e,0) (paw-1.0.0-16e,0) (paw-1.0.0-8e,0) (GBRV-v1-16e$^{[+]}$,0) (VASP-paw$_{2012}$,0) (VASP-paw$_{2000}$,0)  (WC-rrkjus-0.3.0-16e,2.796) (PBESOL-rrkjus-0.2.1-16e,2.792) };
\end{axis}
\begin{axis}[stack plots=y, ymin=2.76001, ymax=2.925, axis y discontinuity=crunch,legend pos=north west]	
    \pgfplotsset{ticks=none}   				
    \addplot+[yellow!84!black!,ultra thick, no marks] coordinates	
        {(0,2.857) (1,2.857)} ;
    \legend{Expt.~\cite{Basinski}}
\end{axis}
\end{tikzpicture}
\end{tabular}
\caption {Equilibrium lattice parameter at 0 K for the different iron pseudopotentials tested in this work. All the data shown here are obtained with the PBE XC functional except for the last three columns on the right, where we have used PBE+U~\cite{anisimov1,anisimov2,anisimov-licht,CococcioniU}, WC~\cite{WC} and a PBEsol~\cite{PBEsol} respectively (here we use a Hubbard $U$ correction with $U=3 eV$). The data come from a Birch-Murnaghan fit, do not include zero-point energy and are compared to all-electron \textsc{WIEN2k}~\cite{WIEN2k}, \protect{\exciting}~\cite{EXCITING} and \textsc{VASP}~\cite{vasp} calculations from Ref.~\onlinecite{Cottenier, Pavone-pp, Cottenier-web} respectively and experiments~\cite{Basinski,Ebert,Nix} 
(horizontal yellow line). The crosshatch dotted column corresponds to the pseudopotential chosen for the production runs. 
}
\label{graph:volume_hist}
\end{center}
\end{figure}
\begin {figure}[] 
\pgfplotsset{title=, width=10.8cm,height=6.2cm,}
\begin{center}
\begin{tabular}{rl}
    \begin{tikzpicture} [baseline,trim axis right]]
    \begin{axis}[
    ybar stacked,
    ylabel={\large $B_0$  (GPa)},
    symbolic x coords={ exciting, WIEN2k,rrkjus-0.2.1-16e, rrkjus-1.0.0-16e, paw-0.2.1-16e, paw-1.0.0-16e, paw-1.0.0-8e, GBRV-v1-16e$^{[+]}$, VASP-paw$_{2012}$, VASP-paw$_{2000}$,  PBE+U-rrkjus-0.2.1-16e, WC-rrkjus-0.3.0-16e, PBESOL-rrkjus-0.2.1-16e},
    xtick=data,
    x tick label style={rotate=45,anchor=east},
    ytick={ 160, 180,200,220,240 }, 
    ymin=141., ymax=252., axis y discontinuity=crunch,
    bar width=5.2 mm,
    every node near coord/.append style={font=\footnotesize},
    nodes near coords 
    ]
\addplot+[ybar, blue!62, draw=blue] plot coordinates { (exciting,0) (WIEN2k,0)  (rrkjus-0.2.1-16e,0)  (rrkjus-1.0.0-16e,172) (paw-0.2.1-16e,194) (paw-1.0.0-16e,200) (paw-1.0.0-8e,201) (GBRV-v1-16e$^{[+]}$,178) (VASP-paw$_{2012}$,193) (VASP-paw$_{2000}$,186) (PBE+U-rrkjus-0.2.1-16e,0) (WC-rrkjus-0.3.0-16e,0) (PBESOL-rrkjus-0.2.1-16e,0)};
\addplot+[ybar, blue!94, postaction={pattern=crosshatch dots}] plot coordinates { (exciting,0) (WIEN2k,0) (rrkjus-0.2.1-16e,200) (rrkjus-1.0.0-16e,0) (paw-0.2.1-16e,0) (paw-1.0.0-16e,0) (paw-1.0.0-8e,0) (GBRV-v1-16e$^{[+]}$,0) (VASP-paw$_{2012}$,0) (VASP-paw$_{2000}$,0) (PBE+U-rrkjus-0.2.1-16e,0) (WC-rrkjus-0.3.0-16e,0) (PBESOL-rrkjus-0.2.1-16e,0) };
\addplot+[ybar, blue!62,postaction={pattern=north east lines}] plot coordinates {  (exciting,188) (WIEN2k,198) (rrkjus-0.2.1-16e,00) (rrkjus-1.0.0-16e,0) (paw-0.2.1-16e,0) (paw-1.0.0-16e,0) (paw-1.0.0-8e,0) (GBRV-v1-16e$^{[+]}$,0) (VASP-paw$_{2012}$,0) (VASP-paw$_{2000}$,0) (PBE+U-rrkjus-0.2.1-16e,0)  (WC-rrkjus-0.3.0-16e,0) (PBESOL-rrkjus-0.2.1-16e,0) };
\addplot+[ybar, blue!40, draw=blue] plot coordinates { (exciting,0) (WIEN2k,0)  (rrkjus-0.2.1-16e,0)  (rrkjus-1.0.0-16e,0) (paw-0.2.1-16e,0) (paw-1.0.0-16e,0) (paw-1.0.0-8e,0) (GBRV-v1-16e$^{[+]}$,0) (VASP-paw$_{2012}$,0) (VASP-paw$_{2000}$,0) (PBE+U-rrkjus-0.2.1-16e,159) (WC-rrkjus-0.3.0-16e,242) (PBESOL-rrkjus-0.2.1-16e,235)};
\end{axis}
\begin{axis}[stack plots=y, ymin=141, ymax=252, axis y discontinuity=crunch,legend pos=north west]	
    \pgfplotsset{ticks=none}   	
    \addplot+[yellow!84!black!,ultra thick, no marks] coordinates
        {(0,170.3) (1,170.3)};
    \legend{Expt.~\cite{Adams}}
\end{axis}
\end{tikzpicture}
\end{tabular}
\caption {Equilibrium bulk moduli at 0 K for the different iron pseudopotentials tested in this work. All the data shown here are obtained with the PBE XC functional except for the last three columns on the right, where we have used PBE+U~\cite{anisimov1,anisimov2,anisimov-licht,CococcioniU}, WC~\cite{WC} and a PBEsol~\cite{PBEsol} respectively (we use here a Hubbard $U$ correction with $U=3 eV$). The data come from a Birch-Murnaghan fit, do not include zero-point energy and are compared to all-electron \textsc{WIEN2k}~\cite{WIEN2k}, \protect{\exciting}~\cite{EXCITING} and \textsc{VASP}~\cite{vasp} calculations from Ref.~\onlinecite{Cottenier, Pavone-pp, Cottenier-web} respectively and experiments~\cite{Adams}  (horizontal yellow line). 
The crosshatch dotted column corresponds to the pseudopotential chosen for the production runs. 
}
\label{graph:bulk_hist}
\end{center}
\end{figure}
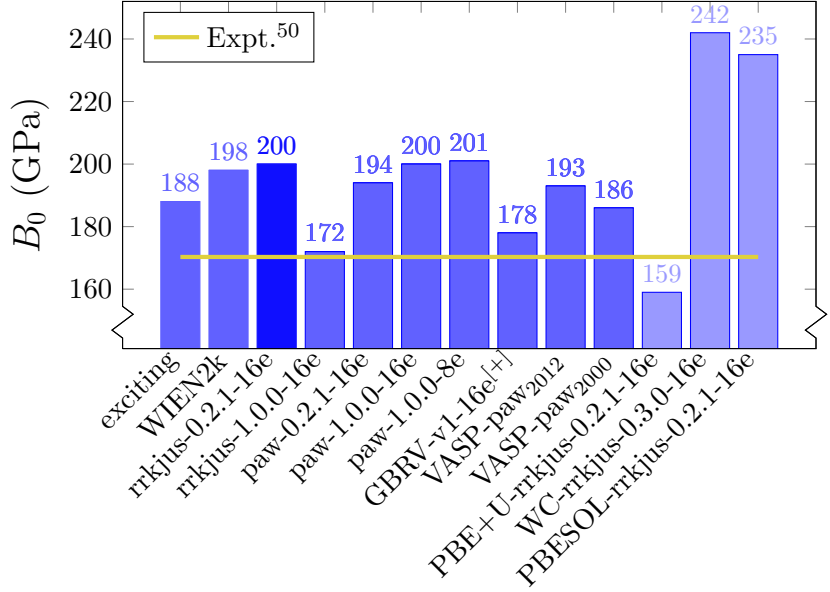
\begin{figure}
\centering
  \includegraphics[trim=0mm 0mm 0mm 0mm, clip, width=0.52\textwidth]{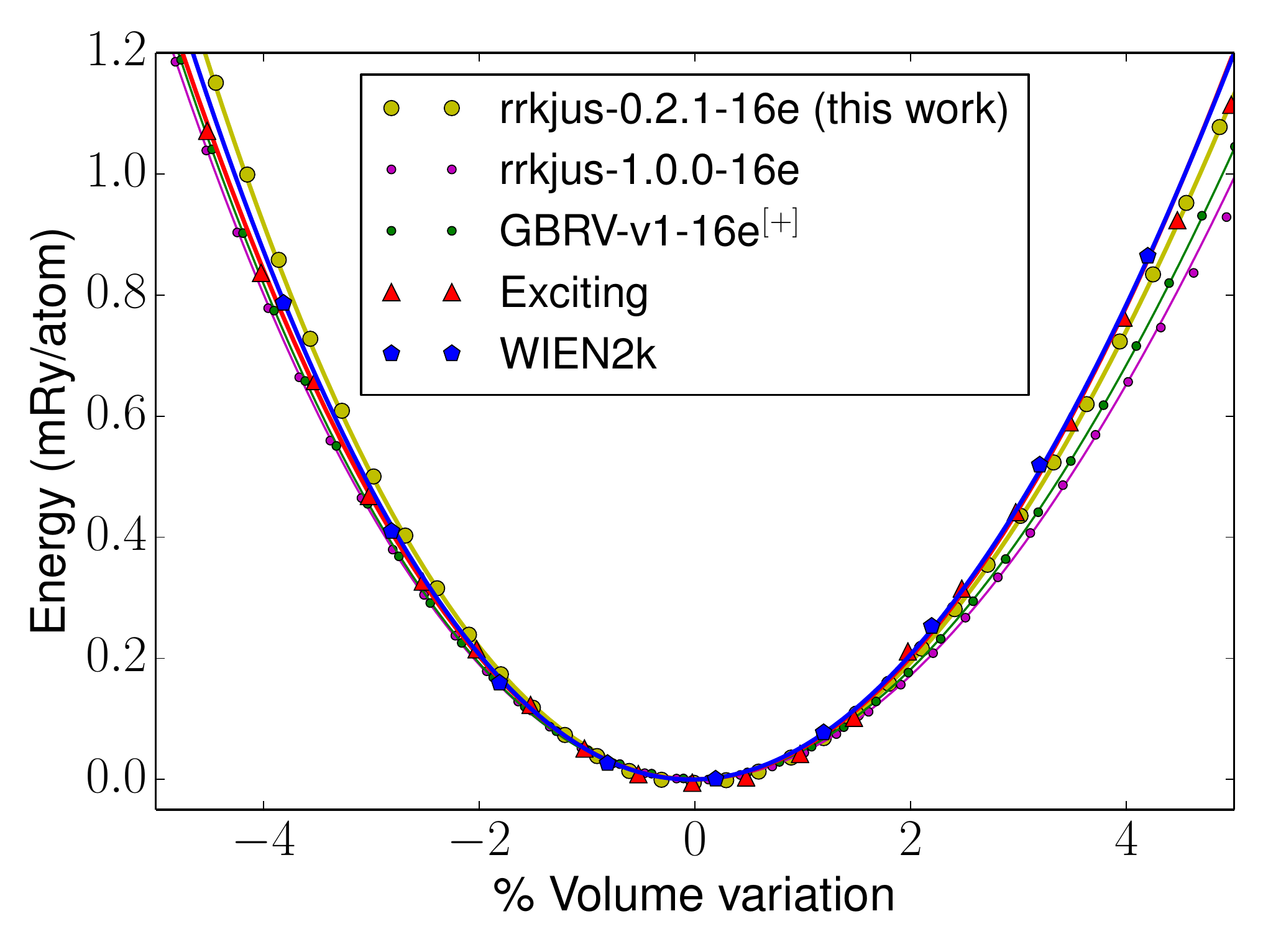} 
  \includegraphics[trim=0mm 0mm 0mm 0mm, clip, width=0.52\textwidth]{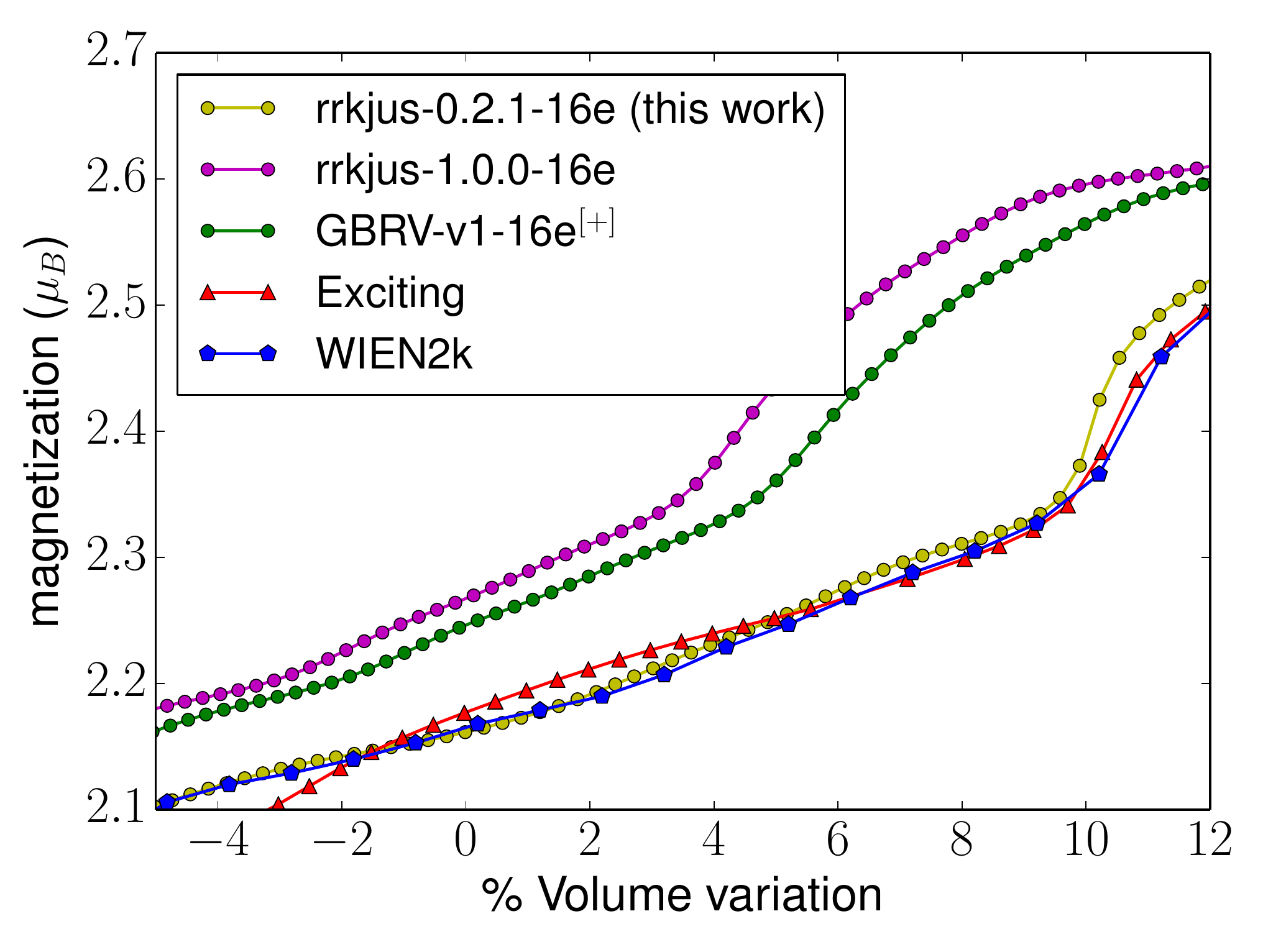} 
  \caption{(Top panel) Equation of state as a function of percent volume change with respect to the theoretical equilibrium configurations for three of the selected PBE pseudopotentials considered in this work (circles). 
  The yellow circles best match the all-electron \textsc{WIEN2k}~\cite{WIEN2k} (pentagons) and \protect{\exciting}~\cite{EXCITING} (triangles) results from Ref.~\onlinecite{Cottenier-pp, Pavone-pp} and correspond to the \texttt{rrkjus-0.2.1-16e} pseudopotential used in this work. 
  Continuous lines are the best fit of the Birch-Murnaghan equation.
(Bottom panel) Total magnetization as a function of percent volume change. The soft magnetic transition discussed in the text is visible as a clear change in the average slope of the different curves.  }
  \label{fig:BM_eos}
\end{figure}

\begin{figure}
\begin{minipage}{\columnwidth}
  \includegraphics[trim=0mm 0mm 0mm 0mm, clip, width=0.44\textwidth]{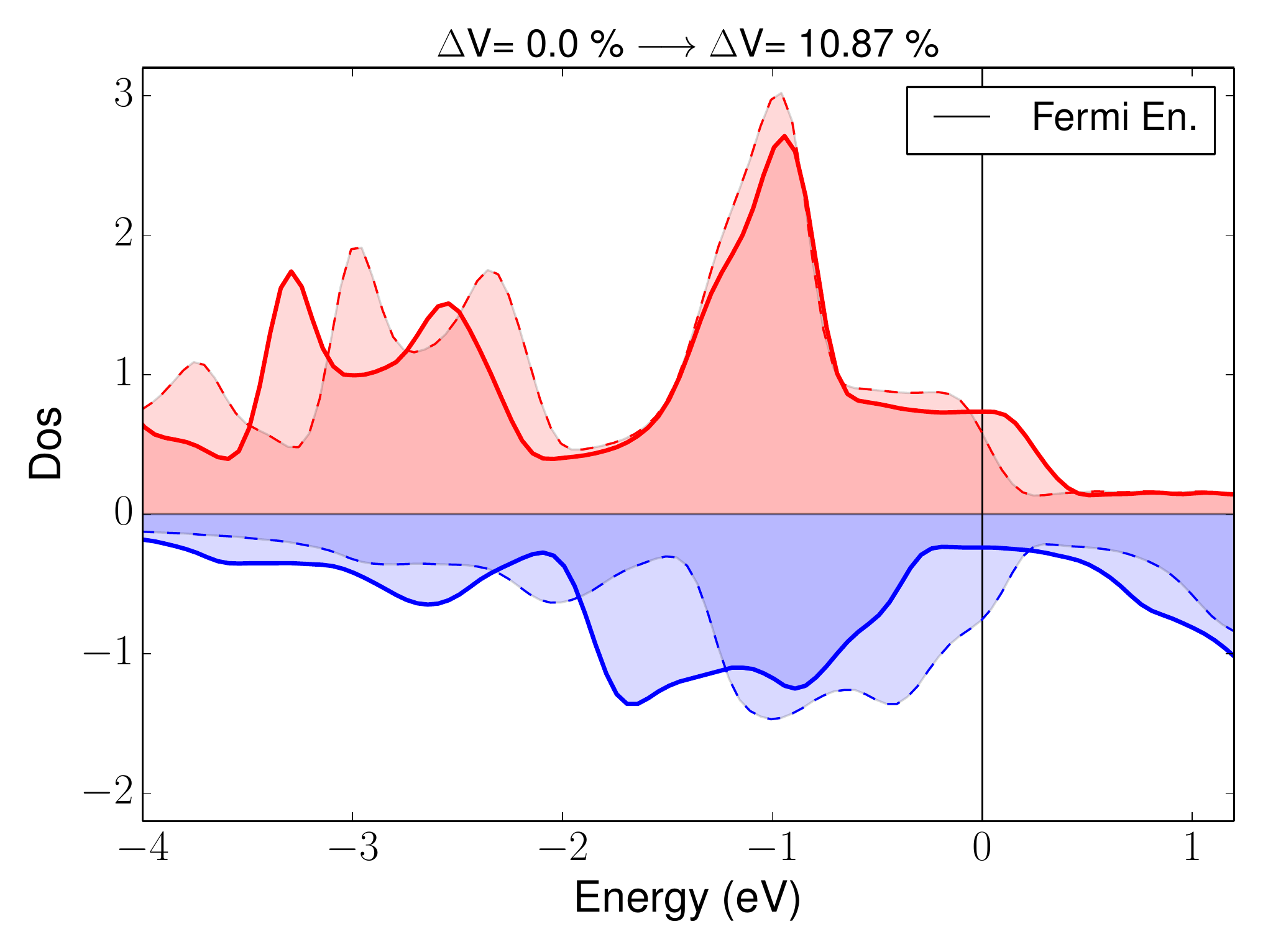} 
  \footnotemark[1]
  \end{minipage}
\begin{minipage}{\columnwidth}
  \includegraphics[trim=0mm 0mm 0mm 0mm, clip, width=0.44\textwidth]{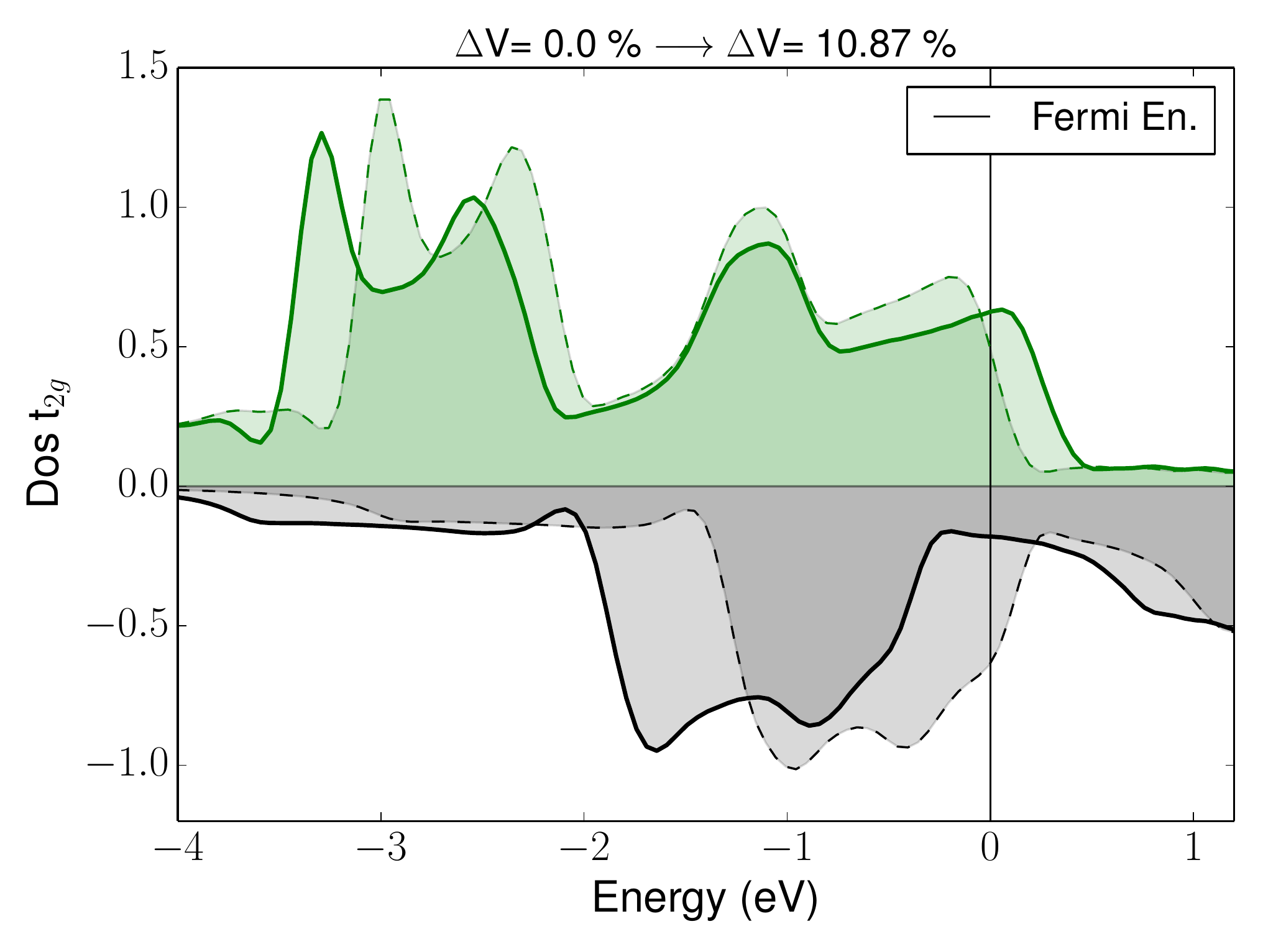} 
  \footnotemark[2]
  \end{minipage}
  \caption{ [a] DOS of majority/minority (red/blue) spin channels at the equilibrium (solid line) and $\Delta V \approx$11\% (dashed line) where, for the pseudopotential used in production run, 
  the magnetic transition takes place. [b] The contribution of the $t_{2g}$ electrons to the majority/minority DOS (green/black) is also reported. 
  To obtain a smooth DOS, a non-self-consistent calculation with an offset $60\times60\times60$ Monkhorst-Pack $k$-mesh is performed on top of a scf loop. }
  \label{fig:PDOS}
\end{figure}
These values in Figs.~\ref{graph:volume_hist},~\ref{graph:bulk_hist} are obtained from a Birch-Murnaghan fit of calculated $E(V)$  data points. 
Interestingly, we have found that the volume range of validity for fitting a Birch-Murnaghan curve is limited on the expansion side due to anomalies in the $E(V)$ curve and its derivatives. These anomalies, also 
reported for all-electron and other calculation methods in Ref.~\onlinecite{EOSpalla}, are more clearly visible as ``shoulders'' in the $M(V)$ behavior (see Fig.~\ref{fig:BM_eos}) and, as visible from Fig.~\ref{fig:PDOS},  
can be associated to a smooth magnetic transition from a low to high spin state due to the splitting of the majority and minority spin $t_{2g}$ electrons upon increasing the volume. 
However, for the pseudopotential chosen here, the expanded volumes at which this anomaly is observed (above 9\%~\footnote{For the GBRV and rrkjus-1.0.0 pseudopotentials reported in Fig.~\ref{fig:BM_eos} instead, the anomaly 
starts around +4/5\% of their equilibrium volume and the magnetization is systematically overestimated if compared to all-electron data.}) are far beyond the theoretical thermal expansion of the system in the thermodynamic region 
considered in this work, thus enabling us to fit the energy surface with volume expansions up to $\sim$9\% still using a standard Birch-Murnaghan equation.

\section{Results}
\label{sec:results}
In this section we present results for selected thermodynamic quantities and for the three strain deformations (hydrostatic or volumetric, tetragonal, and trigonal). Each deformation determines uniquely one of the three elastic constants: $B$ (bulk modulus), $C_{11}$ and $C_{44}$ respectively.

\subsection{$B$ -- Volumetric strain}
\begin{figure}
\centering
  \includegraphics[trim=0mm 0mm 0mm 0mm, clip, width=0.46\textwidth]{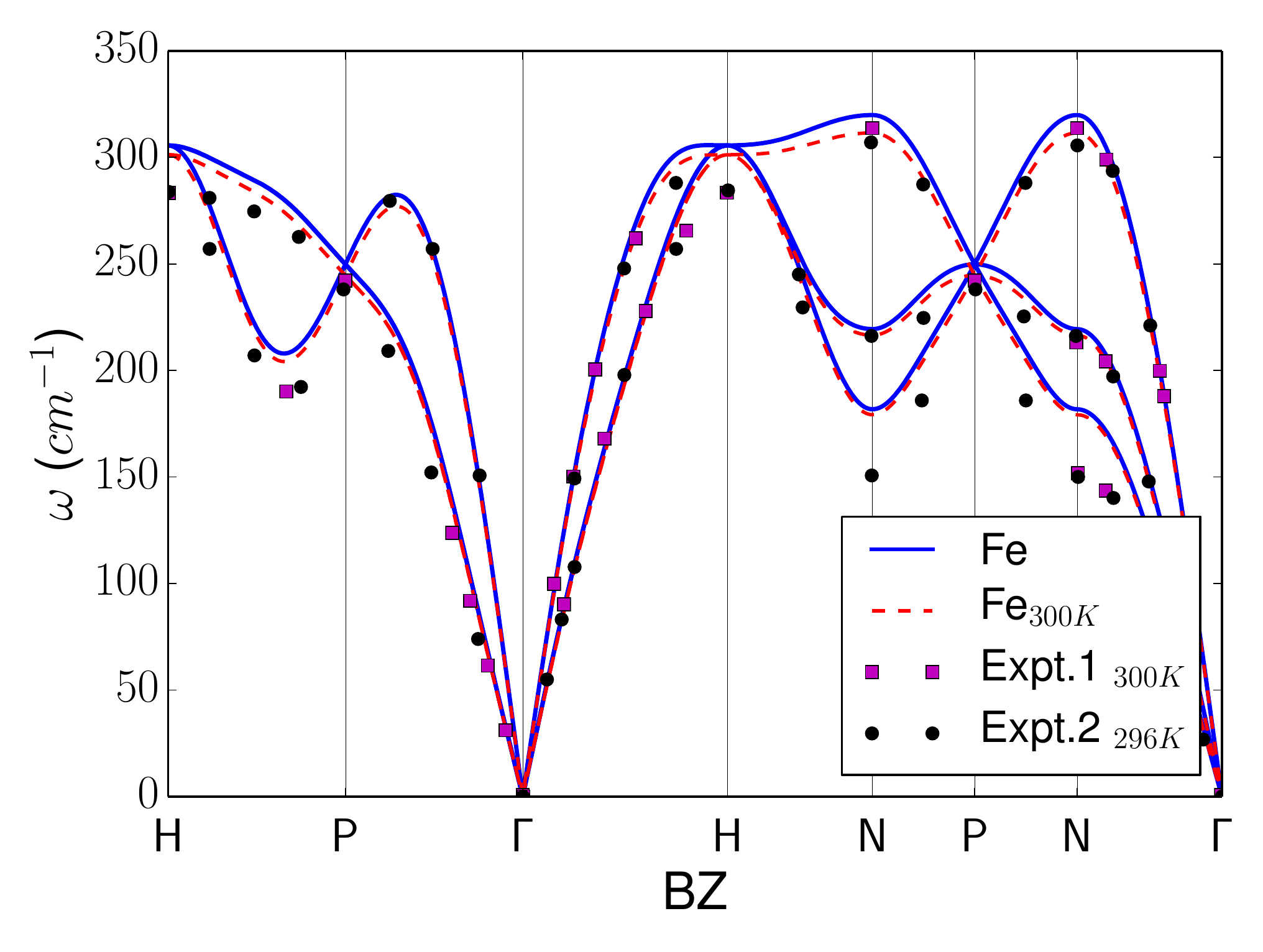} 
  \includegraphics[trim=0mm 0mm 0mm 0mm, clip, width=0.46\textwidth]{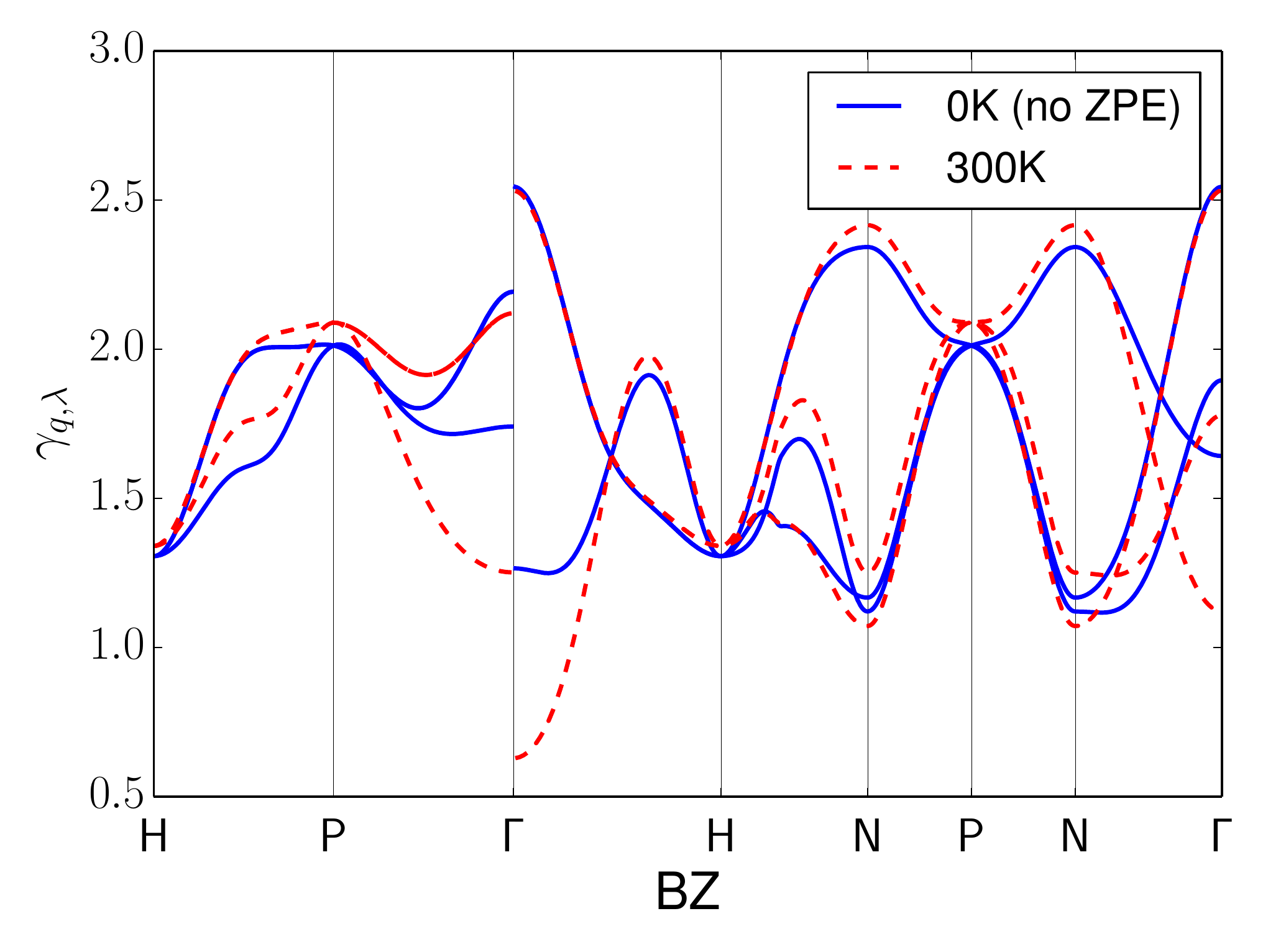} 
  \caption{(Left panel) Phonon dispersions along high-symmetry directions in the BZ calculated at the theoretical electronic equilibrium volume (blue solid line) and at the quasi-harmonic 
  theoretical equilibrium volume at 300~K (red dashed line). The results (see also  Fig.~1 in Supplemental Material~\footnote{See EPAPS Document No.[]} and Ref.~\onlinecite{shobhana} or Ref.~\onlinecite{DalCorsodeGir} for comparison with previous theoretical data) 
  are compared to experimental data at room temperature from Ref.~\onlinecite{Braden} (Expt.1 -- squares) and Ref.~\onlinecite{Brockhouse} (Expt.2 -- circles).
  (Right panel) Gr\"{u}neisen parameters calculated along the same path in the BZ  and the same equilibrium volumes used for the phonon dispersion (blue solid line for the 0~K case and red dashed line for the 
  300~K case). The Gr\"{u}neisen parameters are obtained computing the first derivative with respect to the volume of a cubic fit of the phonon frequencies.  }
  \label{fig:phonon}
\end{figure}

The volumetric deformation $\boldsymbol{\varepsilon}^{(1)}$ can be described by a
single parameter $\varepsilon_a$, namely, the strain of the cubic lattice parameter (see Tab.~\ref{tab:deform}). Thus, the lattice spacing
is defined as:
\begin{align}
  a &= a_0 (1+\varepsilon_a) \label{eq:strain-a}
\end{align}
where $a_0$ is the theoretical equilibrium lattice parameter without zero-point contribution~(see Tab.~\ref{tab:C0Kab}).
The static part of the Helmholtz free energy of Sec.~\ref{sec:framework} is obtained by fitting a
Birch-Murnaghan equation of state~\cite{birch-murnaghan} to a series of well converged total energy values
calculated on a one dimensional regular grid with $\varepsilon_a$ going from $-$0.02\, to $+$0.03 in steps
of 0.001. The resulting static contribution to the bulk modulus is reported in Tab.~\ref{tab:C0Kab}.
The vibrational contribution, on the other hand, has been calculated on a coarser grid via integration of the phonon dispersions as from Eq.~\ref{eq:QHA} 
(examples for the calculated phonon dispersion and resulting Gr\"{u}neisen parameters can be found in Fig.~\ref{fig:phonon}), with $\varepsilon_a$ ranging from $-$0.012 to $+$0.020 in increments of 
0.004 and fitted with a second-order polynomial as a function
of the strain parameter $\varepsilon_a$. The stability of the results has been checked against a fit with
lower and higher order polynomials (see Supplemental Material~\footnote{See EPAPS Document No.[]} ).

\begin{figure}
\centering
  \includegraphics[trim=5mm 5mm 5mm 5mm, clip, width=0.52\textwidth]{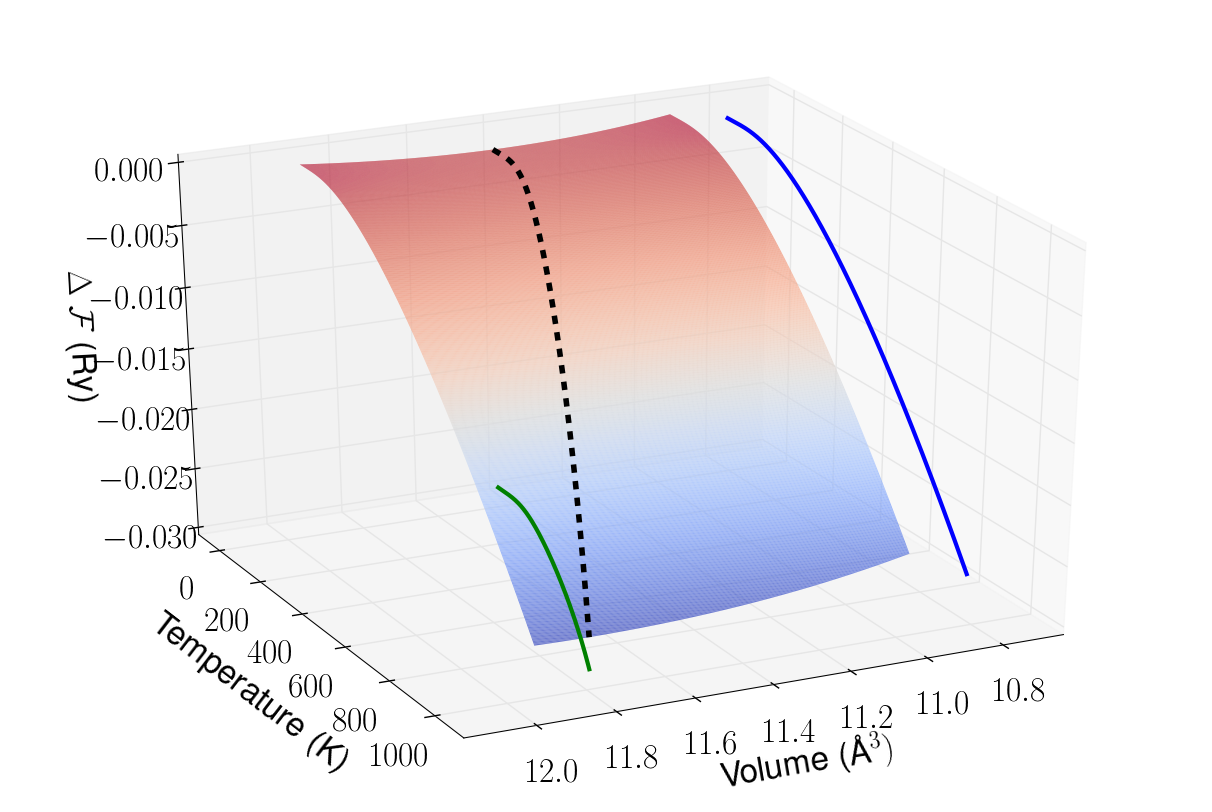} 
  \includegraphics[trim=0mm 0mm 0mm 0mm, clip, width=0.52\textwidth]{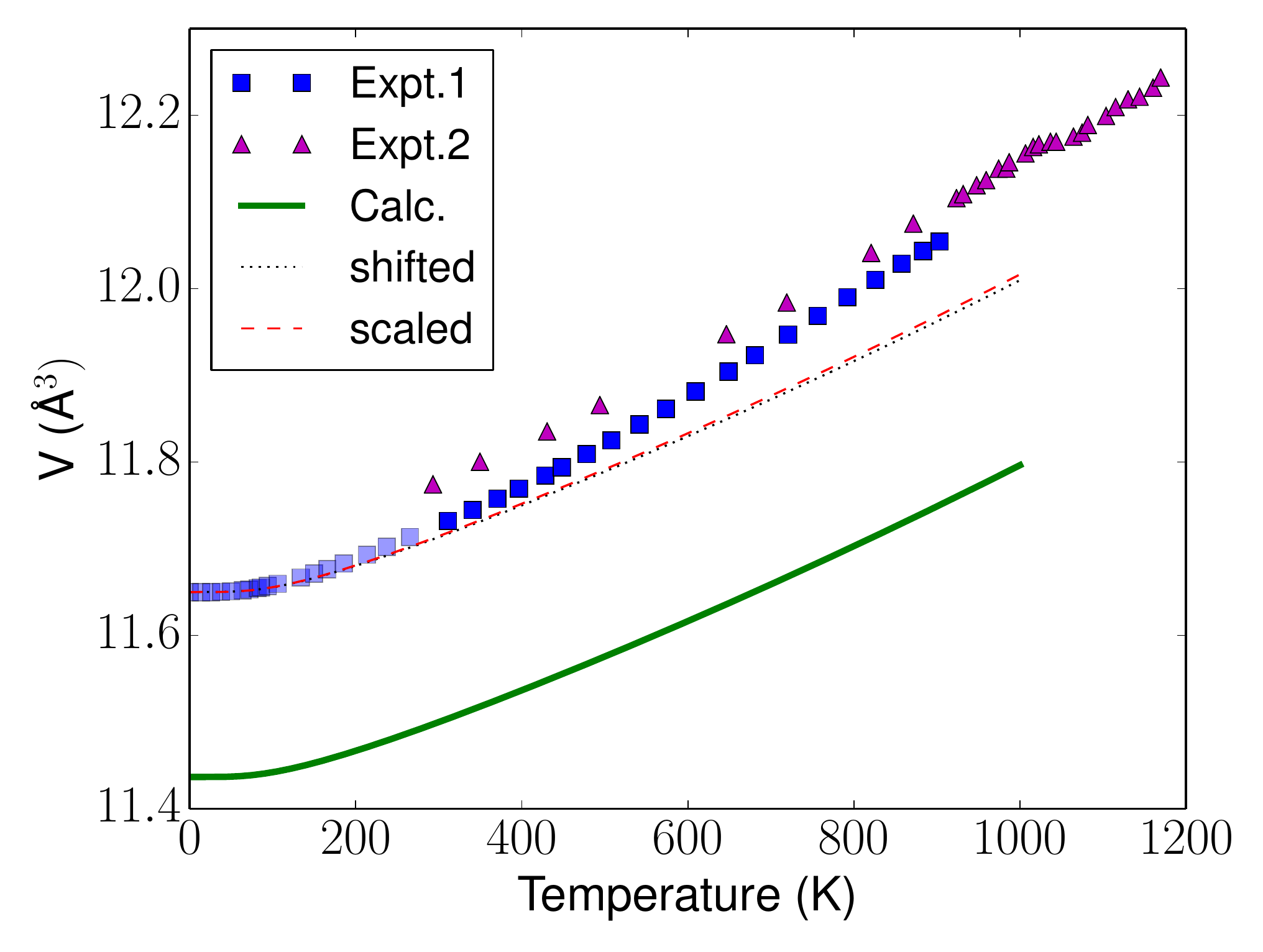} 
  \caption{(Top panel) Free-energy landscape of cubic BCC iron as a function of volume $V$ and temperature $T$.
  The dashed black line corresponds to the set of points that minimize the free-energy surface at each
  temperature. The continuous green and blue lines are the projections of the black dashed line in the $T$-$V$
  and $F$-$T$ planes, thus describing the volumetric thermal expansion and the zero-pressure free energy as a
  function of $T$. 
(Bottom panel) Volumetric thermal expansion (green solid line) compared to experimental data from Ref.~\onlinecite{Basinski} (Expt.1 -- blue squares, note that below room temperature the data 
are extrapolated according to the thermal expansion coefficient of Ref.~\onlinecite{Nix}) and Ref.~\onlinecite{Ridley} (Expt.2 -- magenta triangles).
As a guide to the eye, we also report in all plots shifted and scaled quantities. The former are rigidly translated on the vertical axis while the latter are multiplied by a constant factor to match 
the experimental 0~K value.}
  \label{fig:free_iso}
\end{figure}

\begin{figure}
\centering
  \includegraphics[trim=0mm 0mm 0mm 0mm, clip, width=0.44\textwidth]{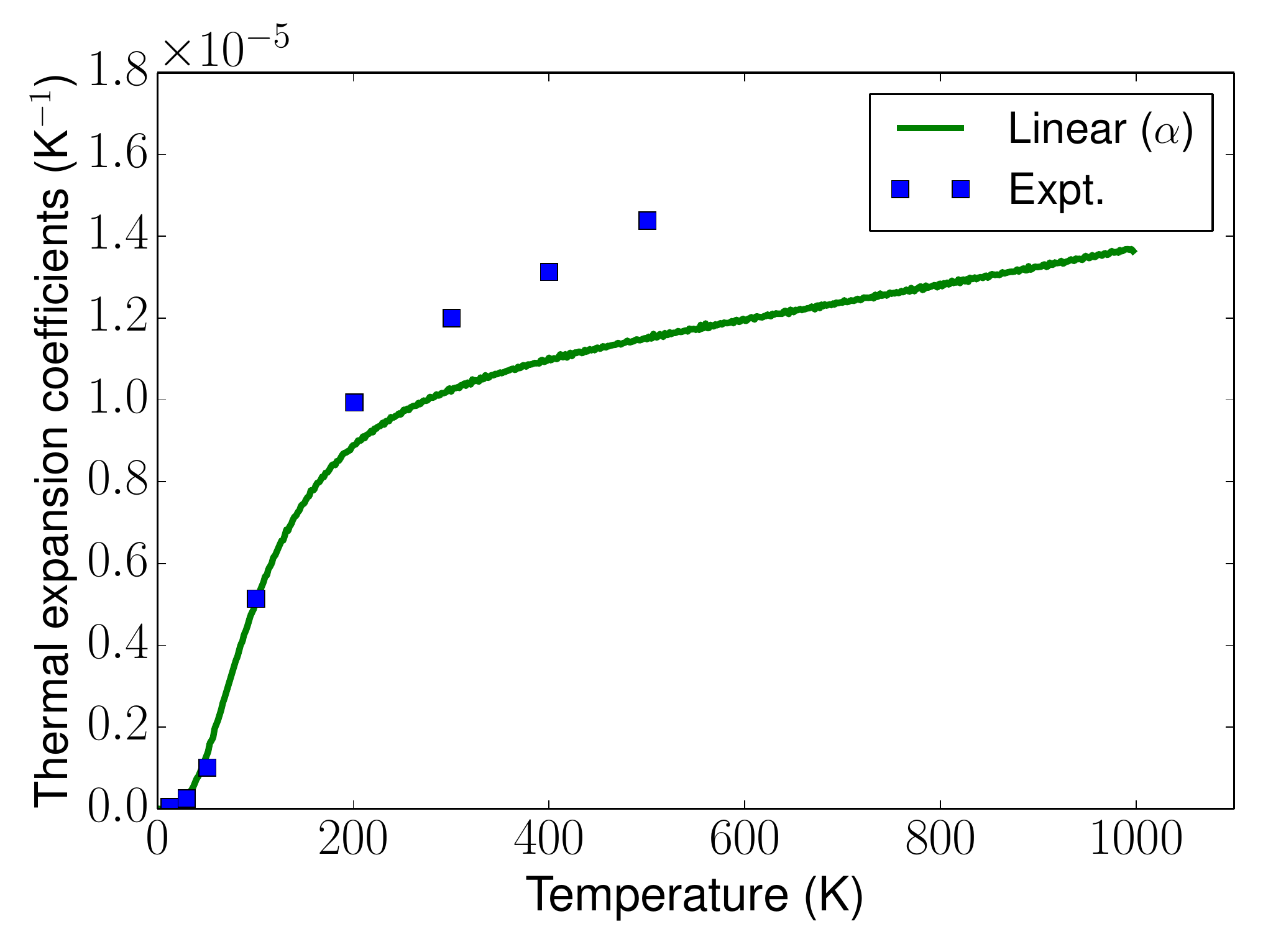} 
  \includegraphics[trim=0mm 0mm 0mm 0mm, clip, width=0.42\textwidth]{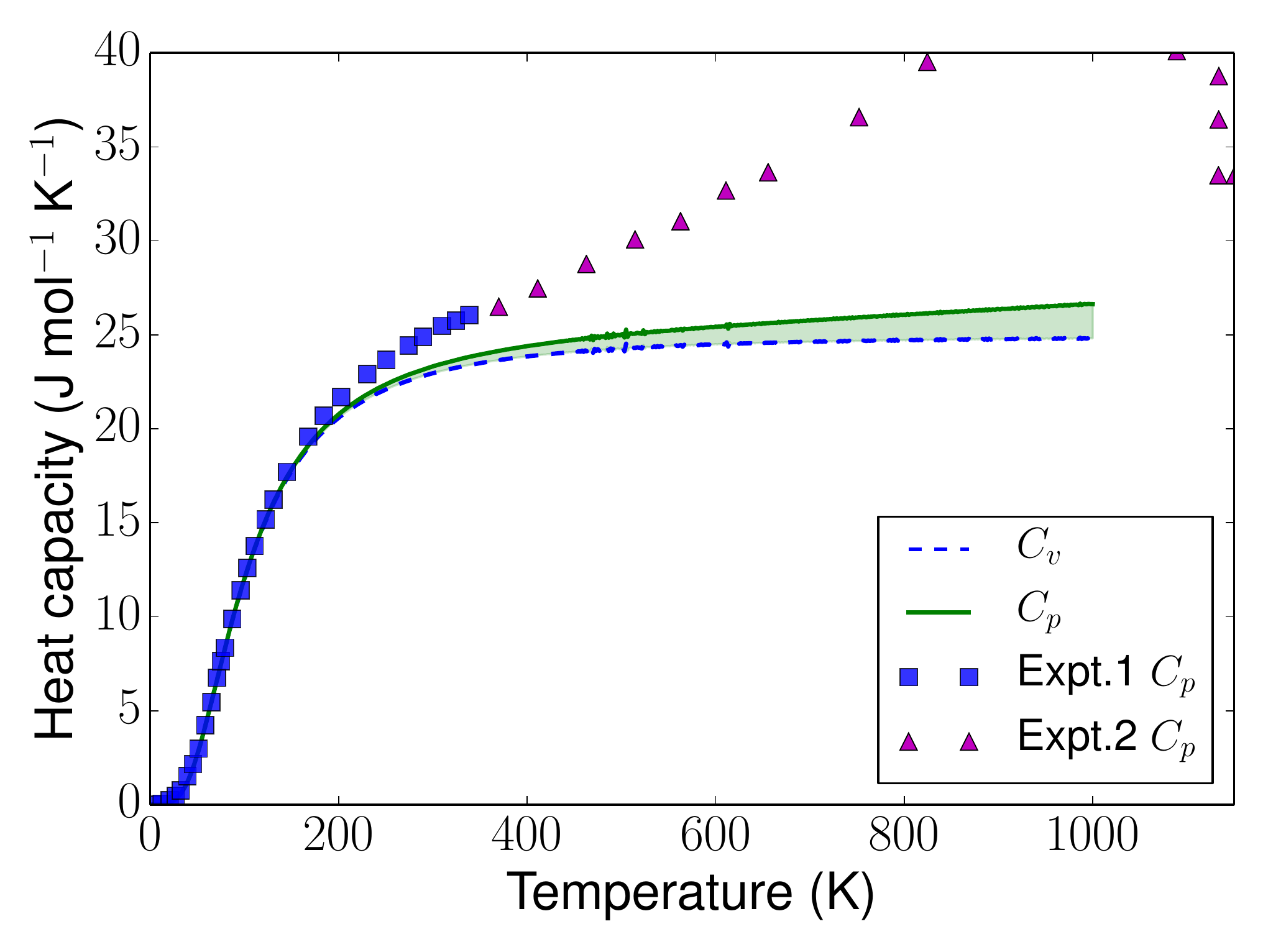} 
  \caption{(Left panel) Linear (green solid line) coefficient of thermal expansion compared to experimental data from Ref.~\onlinecite{Grigoriev} (squares). (Right panel) Specific heat at constant pressure 
  (green solid line), at constant volume (blue dashed line), and compared to experimental data from Ref.~\onlinecite{Desai} (Expt.1 -- squares) and from Ref.~\onlinecite{Wallace2} (Expt.2 -- triangles) .}
  \label{fig:thermodynamics}
\end{figure}

\begin{figure}
\centering
  \includegraphics[trim=0mm 0mm 0mm 0mm, clip, width=0.52\textwidth]{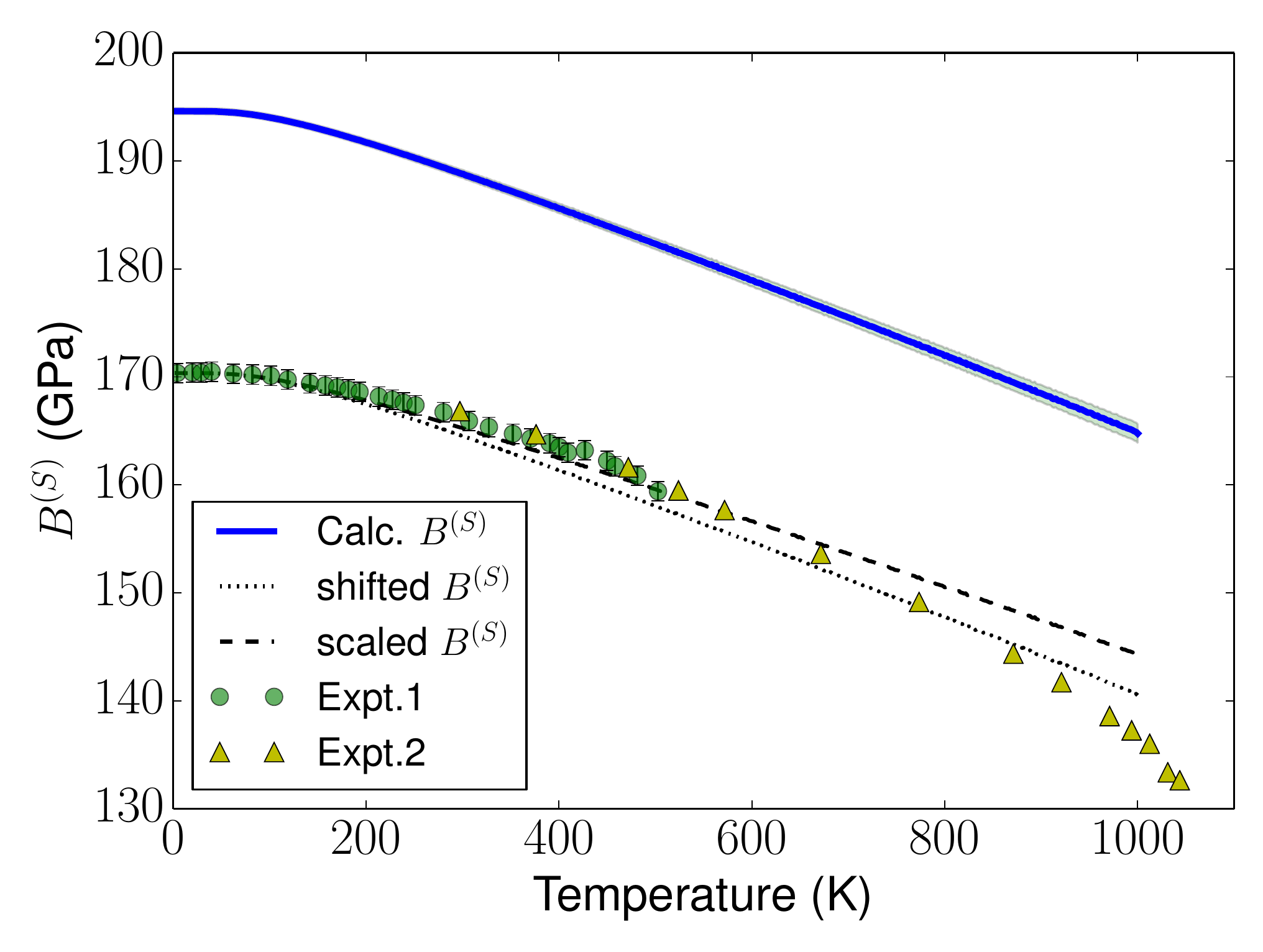} 
  \caption{Adiabatic bulk modulus as a function of $T$ (blue continuous line) calculated along with its confidence interval on the fit (shaded green) and compared to experimental data
  from Ref.~\onlinecite{Adams} (Expt.1 -- green circles) and from Ref.~\onlinecite{Dever} (Expt.2 -- yellow triangles).
  As a guide to the eye, we also plot the bulk modulus rigidly shifted (dotted line) and scaled (dashed line) to match the experimental 0~K value.}
  \label{fig:Bulk_T}
\end{figure}

The free energy is then obtained as an analytical function of $\varepsilon_a$ and $T$ and is shown in
Fig.~\ref{fig:free_iso}. We then determined the thermal expansion (Fig.~\ref{fig:free_iso}), the thermal expansion coefficients (Fig.~\ref{fig:thermodynamics}), 
the heat capacity (Fig.~\ref{fig:thermodynamics}) and the isothermal bulk modulus
$B^{(T)}(T)$ from the analytic second derivative of the free energy as in Eq.~\ref{eq:El-def}. The
adiabatic correction of Eq.~\ref{eq:Isotoadia} is then used  to compute the adiabatic bulk modulus
$B^{(S)}(T)$. Results are reported in Fig.~\ref{fig:Bulk_T} and compared to experimental data from
Refs.~\onlinecite{Adams,Dever}.
The agreement between experiments and calculations in the thermal behavior of the bulk modulus is remarkable,
especially below the Debye temperature ($\Theta_D \simeq$~500~K). Above $\Theta_D$, the small deviation from 
experiments can be ascribed to magnetic fluctuations~\cite{Hasegawa,Dever,Fultz,Rusanu} that become increasingly
important approaching the Curie temperature (1043~K), plus minor contributions from anharmonic effects
(beyond quasi-harmonic) and from the electronic entropy. At 1000~K, the softening of the calculated $B^{(S)}$
is nearly 15\% with respect 0~K. The calculated magnetic moment increases from 2.17~$\mu_B$ per atom (2.22 $\mu_B$ from experiments~\cite{Kittel}) at the 0~K equilibrium volume 
to 2.27~$\mu_B$ at the 1000~K equilibrium volume. Obviously, transverse magnetic fluctuations are neglected in these calculations, and we postpone to Sec.~\ref{sec:discussion} the
discussion on the mismatch between experiments and calculations in absolute values.

\subsection{$C_{11}$, $C_{44}$ -- Tetragonal and Trigonal strains}
The Helmholtz free energy $F$ depends upon two strain parameters: the isotropic lattice strain $\varepsilon_a$ and a second strain parameter $\varepsilon_c$ or $\varepsilon_d$ according to the deformation considered
(see Tab.~\ref{tab:deform}).

The tensor $\boldsymbol{\varepsilon}^{(2)}$ is associated to a continuous tetragonal deformation that
stretches the edge $c$ of the cubic undistorted structure along the $z$ axis while leaving unchanged
the other edges. The relation between the strain $\varepsilon_c$ and the distorted edge $c$ is:
\begin{align} 
  c = a(1+\varepsilon_c). \label{eq:deftet}
\end{align}
The tensor $\boldsymbol{\varepsilon}^{(3)}$ is associated to a continuous trigonal deformation that stretches
the main diagonal $d$ of the undistorted cubic structure along the (111) direction while tilting the
undistorted edges and preserving their length. In this case, the relation between the strain $\varepsilon_d$
and the distorted main diagonal is
\begin{align} 
  d=\sqrt{3}a(1+\varepsilon_d),  \label{eq:deftri}
\end{align}
while the relation with the cosine of the angle between the distorted edges is
\begin{align} 
  cos(\alpha)=\frac{1-\varepsilon_d (2+\varepsilon_d)}{(\varepsilon_d-1)(\varepsilon_d+3)}.
  \label{eq:deftri2}
\end{align}
Both deformations do not conserve the volume per atom. In particular, in the tetragonal one the volume
increases as a function of $\varepsilon_c$, while in the trigonal case, the volume decreases as a function
of $\varepsilon_d$. Alternatively, we could have chosen volume-conserving deformations as in
Ref.~\onlinecite{renata1}, but the advantage of the present scheme is that each deformation determines uniquely
one elastic constant at the time, and enables us to determine easily the confidence interval of each
elastic constant by error-propagation theory.

In the next sub-sections we describe the calculation of the static and vibrational contributions,
separately. The reason is that we want to analyze their contributions to the global energy
landscape separately. This also allows us to sample the two contribution landscapes with two different grids.
Indeed, the static term displays a minimum as a function of the strain parameters and has to be sampled
with a dense grid,  while, on the other hand, the vibrational term is flat, monotonic and can be sampled with
a coarse grid.

\subsubsection{Static contribution}
\begin{figure}
\centering
  \includegraphics[trim=0mm 0mm 0mm 0mm, clip, width=0.52\textwidth]{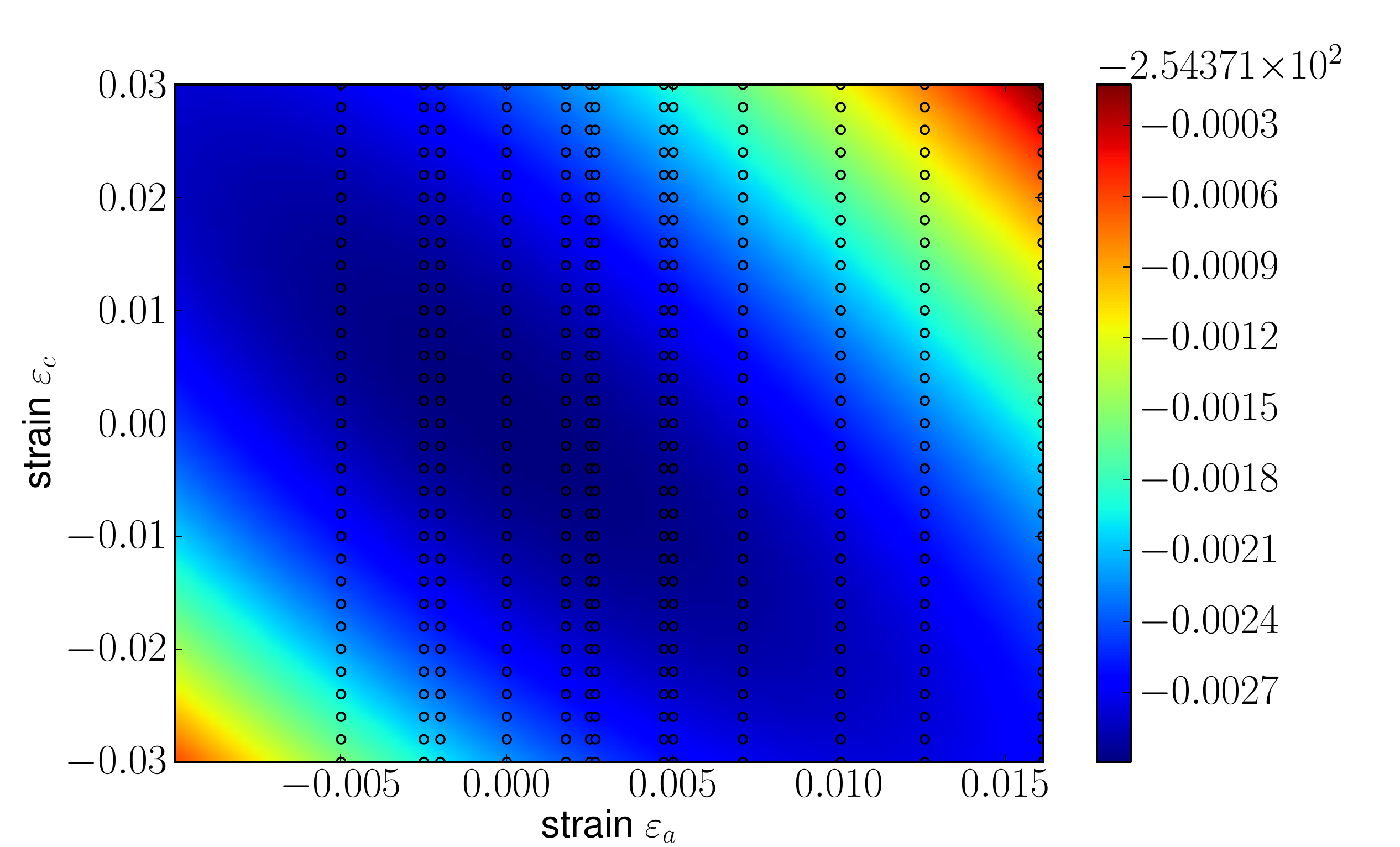} 
  \includegraphics[trim=0mm 0mm 0mm 0mm, clip, width=0.52\textwidth]{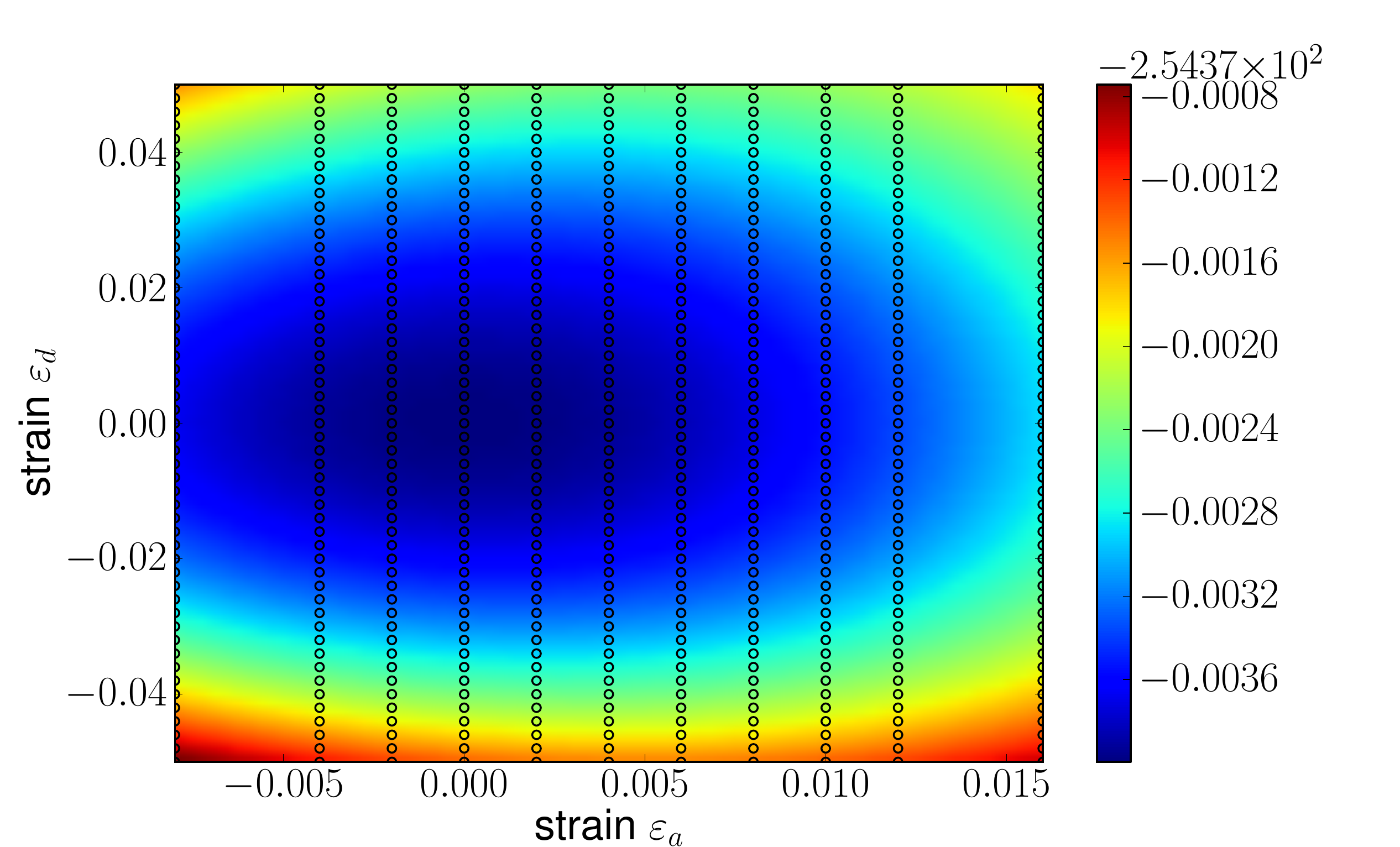} 
  \caption{Static energy landscape of the tetragonal (top panel) and trigonal (bottom panel) distortions
  projected on the $[\varepsilon_a,\varepsilon_{c/d}]$ plane.}
  \label{fig:static-landscape}
\end{figure}

To evaluate the static contribution to the elastic constants, we performed a series of well converged total
energy calculations on a two dimensional discrete grid $[\varepsilon_a,\varepsilon_{c/d}]$ (see Fig.~\ref{fig:static-landscape} for details on the grid). 
The $\varepsilon_a$ grid is asymmetric with respect to zero and with more points in the positive range of the strain parameter, in order to sample accurately 
the values of the static contribution to the free energy also in the thermal expansion range.

The resulting total energies are fitted with a two-dimensional bivariate polynomial up to 5th degree using
a least-square method~\footnote{We used the least squares method routine \texttt{scipy.optimize.leastsq}
which is a wrapper around the Fortran routine \textsl{lmdif} of MINPACK~\cite{More:minpack}.}.

The analysis of the quality of the fit of discrete data points to a two-dimensional energy surface is crucial
to resolve the possible sources of error that could affect our elastic constants and, therefore, for
a reliable comparison with experiments and the wide range of scattered data available in the literature.
Therefore, in addition to the visual inspection of the fit along constant $\varepsilon_{c/d}$ sections, we
evaluated the adjusted coefficient of determination ($R^2$) and the the average absolute error (AAE),
defined as:
\begin{equation}
   \mathrm{AAE} \equiv \frac{1}{N} \sum_{i,j} \left| P_n(\varepsilon_a^{(i)},\varepsilon_{c/d}^{(j)}) - 
   E(\varepsilon_a^{(i)},\varepsilon_{c/d}^{(j)}) \right|,
\end{equation}
where $N$ is the total number of $[\varepsilon_a,\varepsilon_{c/d}]$ discrete values and $P_n$ is the
bivariate polynomial of degree $n$. Thus, $R^2$ is a measure of the quality of the fitting model, i.e. how
well the analytic function approximates the calculated data points. The AAE is a quantitative measure of the
distance of the fitted curve from the calculated points. We found that the AAE decreases by increasing the degree $n$
of the polynomial and $R^2$ approaches unity, as shown in Tab.~\ref{tab:sigma-R2}. According to these
results, in both cases, we considered the 4th-degree polynomial to provide a sufficiently accurate fit
(indeed the AAE is two orders of magnitude smaller than the difference between the highest and the lowest
total-energy data points).

\begin{table}
\centering
\begin{tabular}{ccc@{\hspace{1cm}}cc}
\hline\hline
Order & AAE (Ry) & $R^2$ & AAE (Ry) & $R^2$ \\
      & \multicolumn{2}{c}{tetragonal} & \multicolumn{2}{c}{trigonal} \\
\hline
2 & 1.894\ 10$^{-5}$ & 0.997259 & 9.163\ 10$^{-5}$ & 0.971580\\
3 & 1.997\ 10$^{-6}$ & 0.999975 & 7.369\ 10$^{-6}$ & 0.999819\\
4 & 1.002\ 10$^{-6}$ & 0.999993 & 2.933\ 10$^{-6}$ & 0.999968\\
5 & 9.150\ 10$^{-7}$ & 0.999995 & 1.693\ 10$^{-6}$ & 0.999990\\
\hline\hline
\end{tabular}
\caption{Average absolute error (AAE) and adjusted coefficient of determination ($R^2$) of the
two-dimensional fit of the static energy landscape, for the tetragonal and trigonal deformations,
as a function of the order of the polynomial.}
\label{tab:sigma-R2}
\end{table}
Fig.~\ref{fig:static-landscape} shows a plot of the static energy landscape for both
the tetragonal and trigonal deformations, with the minimum elongated along the
diagonal in the $[\varepsilon_a,\varepsilon_c]$ space or along constant $\varepsilon_d$ in
the $[\varepsilon_a,\varepsilon_d]$ space. 

\subsubsection{Vibrational contribution}
\begin{figure}
  \centering
  \includegraphics[trim=0mm 0mm 0mm 10mm, clip, width=0.52\textwidth]{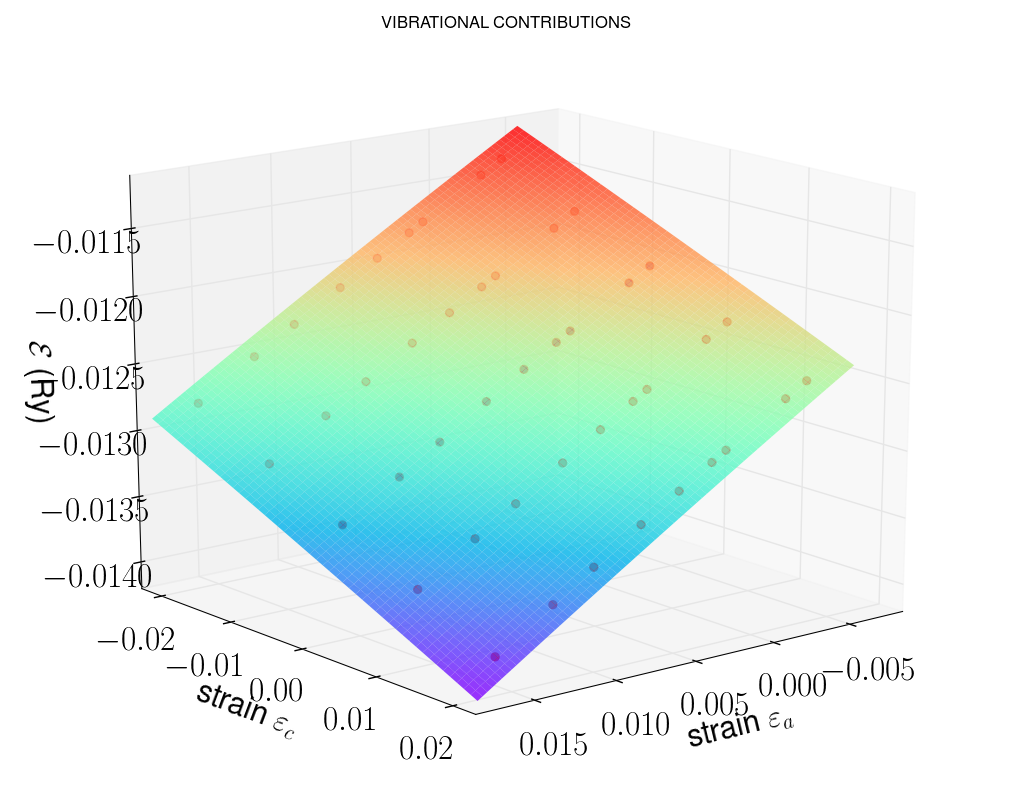} 
  \includegraphics[trim=0mm 0mm 0mm 10mm, clip, width=0.52\textwidth]{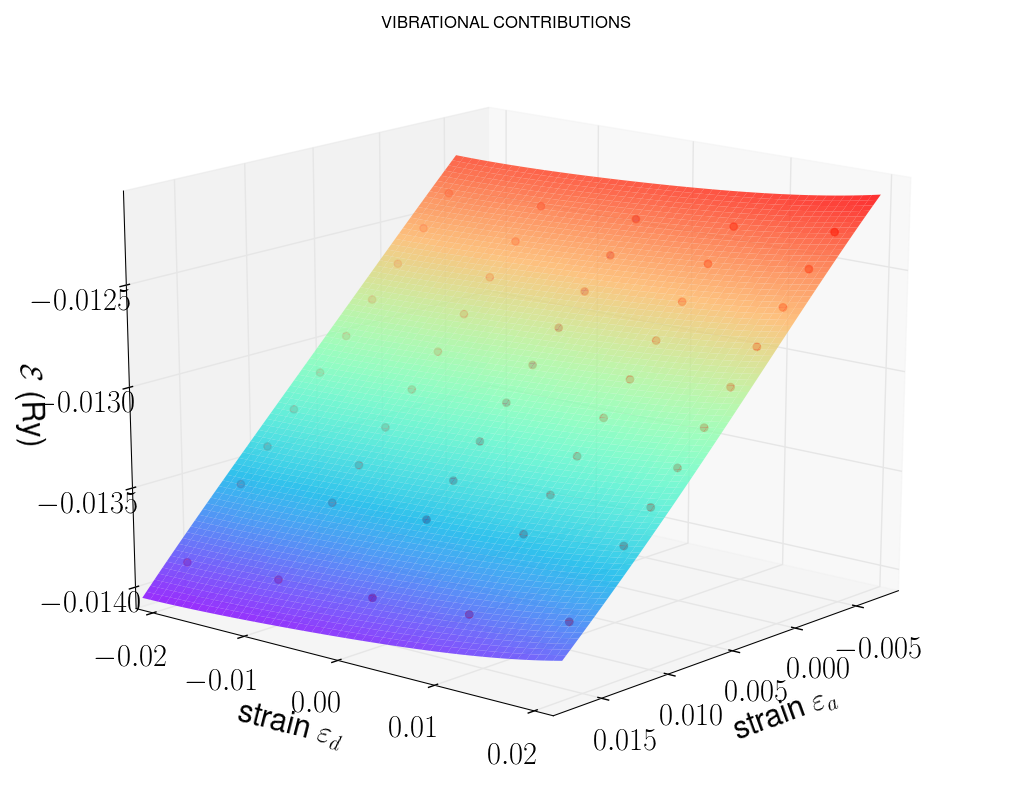} 
  \caption{Vibrational quasi-harmonic contribution to the Helmholtz free energy at $T=750$~K in the
  $[\varepsilon_a,\varepsilon_c]$ space (top panel), $[\varepsilon_a,\varepsilon_d]$ space (bottom panel).
  A 2nd and a 4th order bivariate polynomial are respectively used to fit the tetragonal and trigonal
  data sets. }
  \label{fig:Thermal}
\end{figure}

In order to evaluate the vibrational contributions to the free energy, we performed a series of
linear-response phonon calculations on a two dimensional grid in the space of deformation
scalars $\varepsilon_a$, $\varepsilon_{c/d}$. Since the lattice dynamics calculations are one order of
magnitude more time consuming than the total energy calculations, we used a coarser grid (see Fig.~\ref{fig:Thermal} for details on the grid).

The eigenvalues of each dynamical matrix are Fourier-interpolated in order to obtain smooth and continuous
phonon dispersions. The zero-point energy and the thermal contributions are calculated
by numerical integration over 21$\times$21$\times$21 points in reciprocal space. This is essential to
obtain numerically accurate values of the vibrational contribution.

Like for the case of the static contribution, we determined the best polynomial necessary to fit our data over the
entire temperature range from 0 to 1000~K. Similarly, we used the adjusted $R^2$ and the AAE as
indicators of the quality of the fit. We also checked \textit{a posteriori} the convergence of the
elastic constant curves obtained by fitting to different polynomial degrees. 
In the tetragonal case, a quadratic bivariate polynomial (i.e. 6 parameters) is sufficient to
accurately reproduce the distribution of data points. On the other hand, for the
trigonal deformation, a 4th order bivariate polynomial (i.e. 16 parameters) is needed. 
Our choice of polynomial is dictated by the need to minimize the AAE, maximize $R^2$ and
minimize the confidence interval as a function of temperature (see Supplemental Material~\footnote{See EPAPS Document No.[]} for the stability of the results against other polynomials). As an illustration, we report the
vibrational energy landscape at 750~K for the tetragonal and for the trigonal distortions
(Fig.~\ref{fig:Thermal}).

\subsubsection{Evaluation of the elastic constants}
\begin{figure}
  \centering
  \includegraphics[trim=0mm 0mm 0mm 10mm, clip, width=0.52\textwidth]{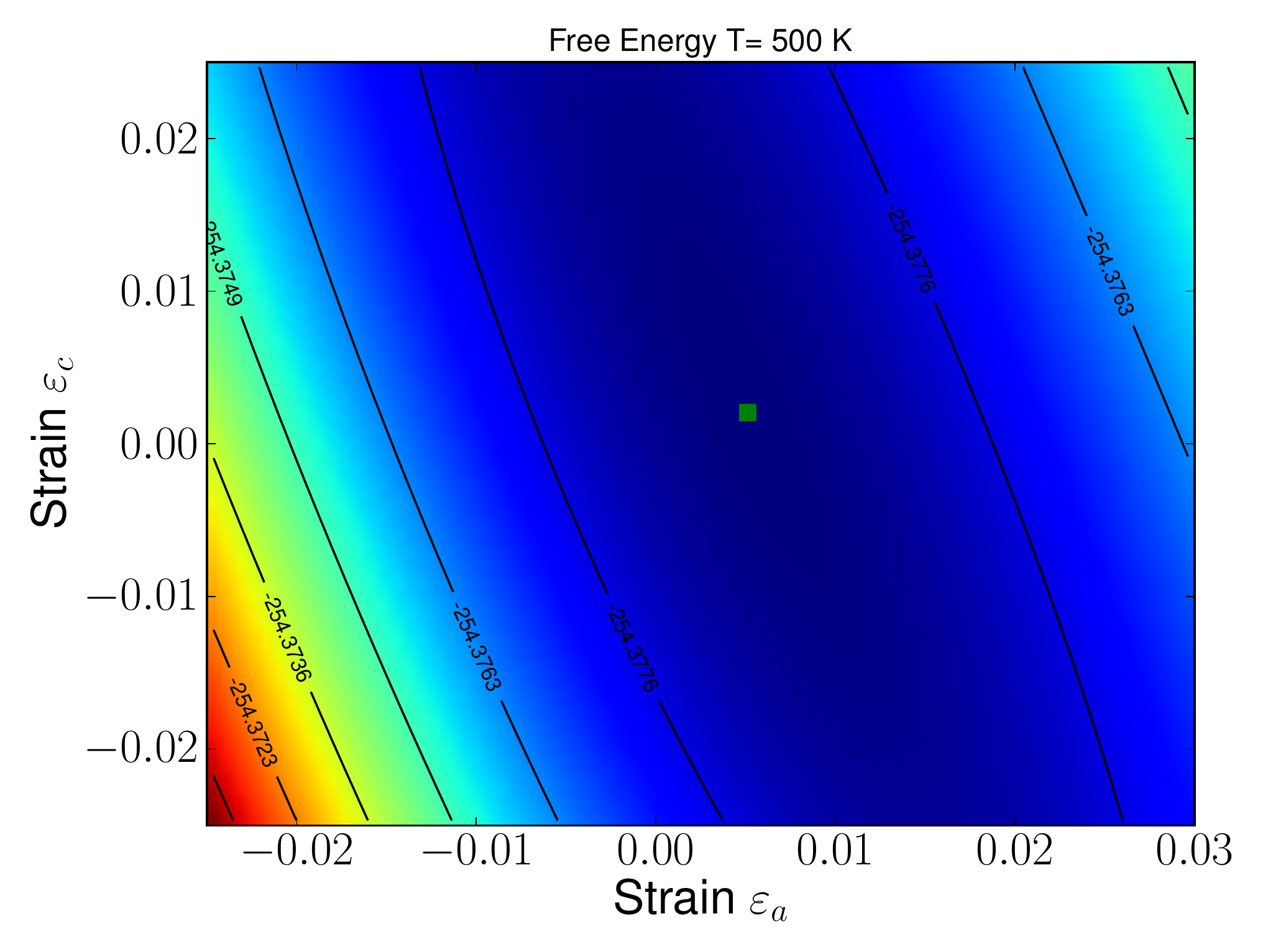} 
  \includegraphics[trim=0mm 0mm 0mm 10mm, clip, width=0.52\textwidth]{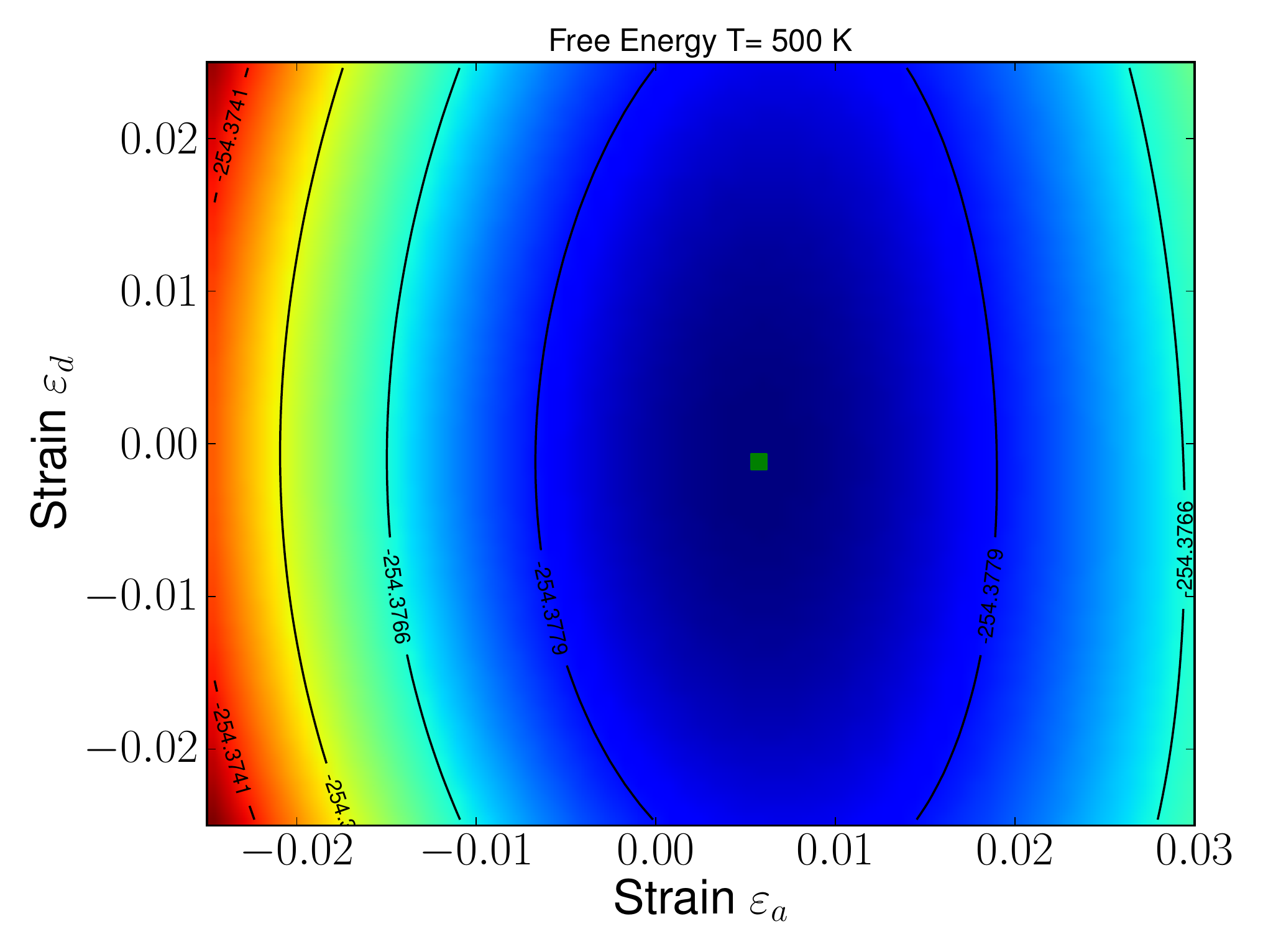} 
  \caption{Free energy landscape of the tetragonal (top panel) and trigonal (bottom panel) deformations
  projected on the $[\varepsilon_a,\varepsilon_d]$ and $[\varepsilon_a,\varepsilon_d]$ plane respectively
  at 500~K. The green squares represent the minima of the energy landscapes.}
  \label{fig:Free}
\end{figure}

Next, we sum the static and vibrational energy landscapes obtained in the previous sections and compute 
the Helmholtz free energy. An example of the resulting landscape at 500~K is 
displayed in Fig.~\ref{fig:Free}. The second derivative with respect to strain can be evaluated analytically 
at the minimum of the free energy as a function of temperature.

Then, in order to understand if the discrepancy between the experimental and calculated elastic constants
could be ascribed to the fitting procedure, we have calculated the confidence interval of $C_{11}$ 
and $C_{44}$. To this end, we have computed the covariance matrix of each best-fit contribution to the 
free energy, defined as:
\begin{align}
  \cov(P) = \sigma_r^2 \left(J^T J\right)^{-1},
\end{align} 
where $P$ is the set of polynomial coefficients, $\sigma_r^2$ is the squared residual and $J$ is the
Jacobian matrix which is provided in output by the least squares routine. The global variance of each
best fit polynomial is then obtained by considering both the diagonal and the off-diagonal elements of 
the covariance matrix $\cov(P)$. Finally, we used error-propagation theory to obtain the confidence
interval of the elastic constants.

\begin{figure}
  \centering
  \includegraphics[trim=0mm 0mm 0mm 0mm, clip, width=0.52\textwidth]{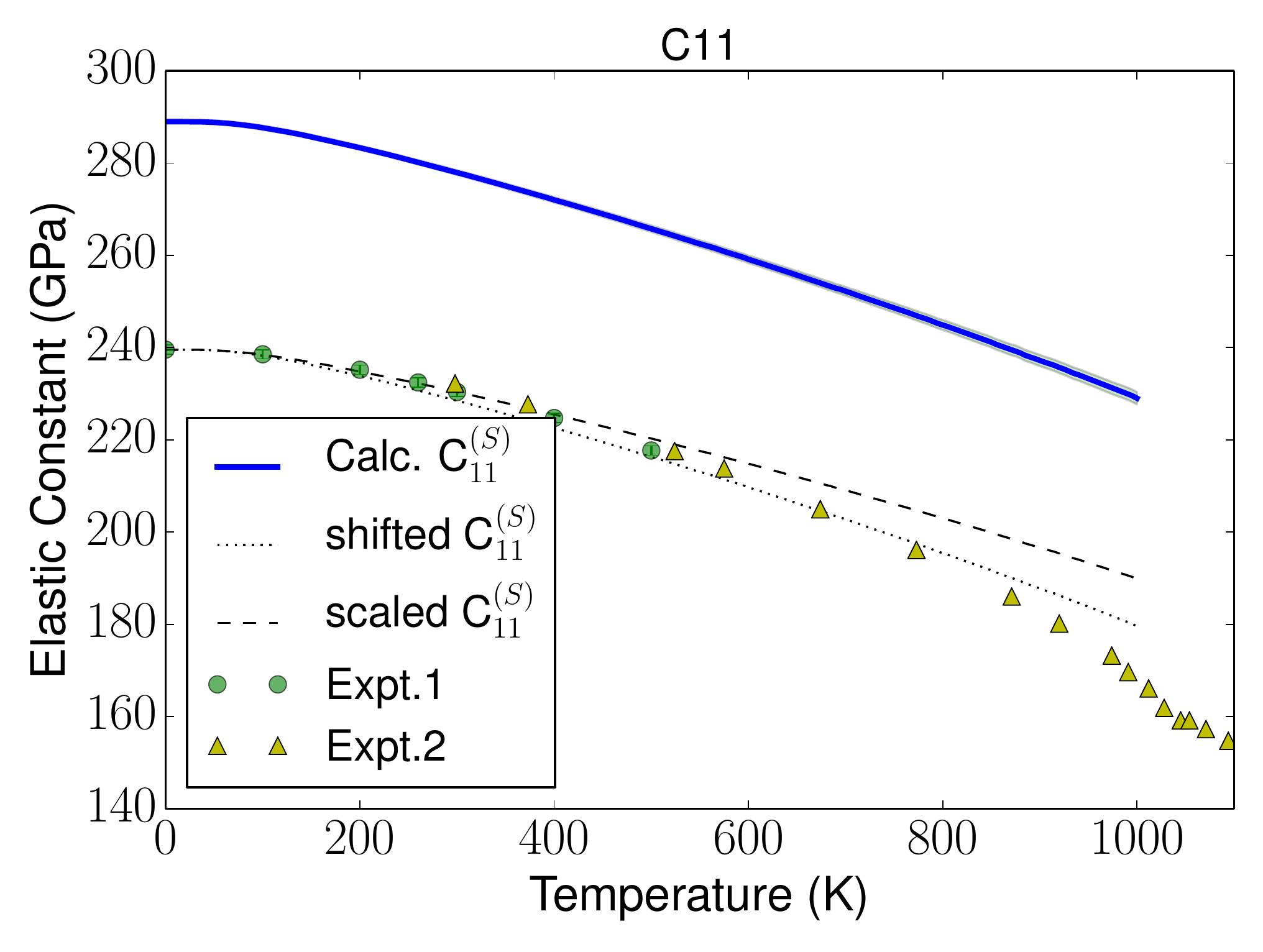} 
  \includegraphics[trim=0mm 0mm 0mm 0mm, clip, width=0.52\textwidth]{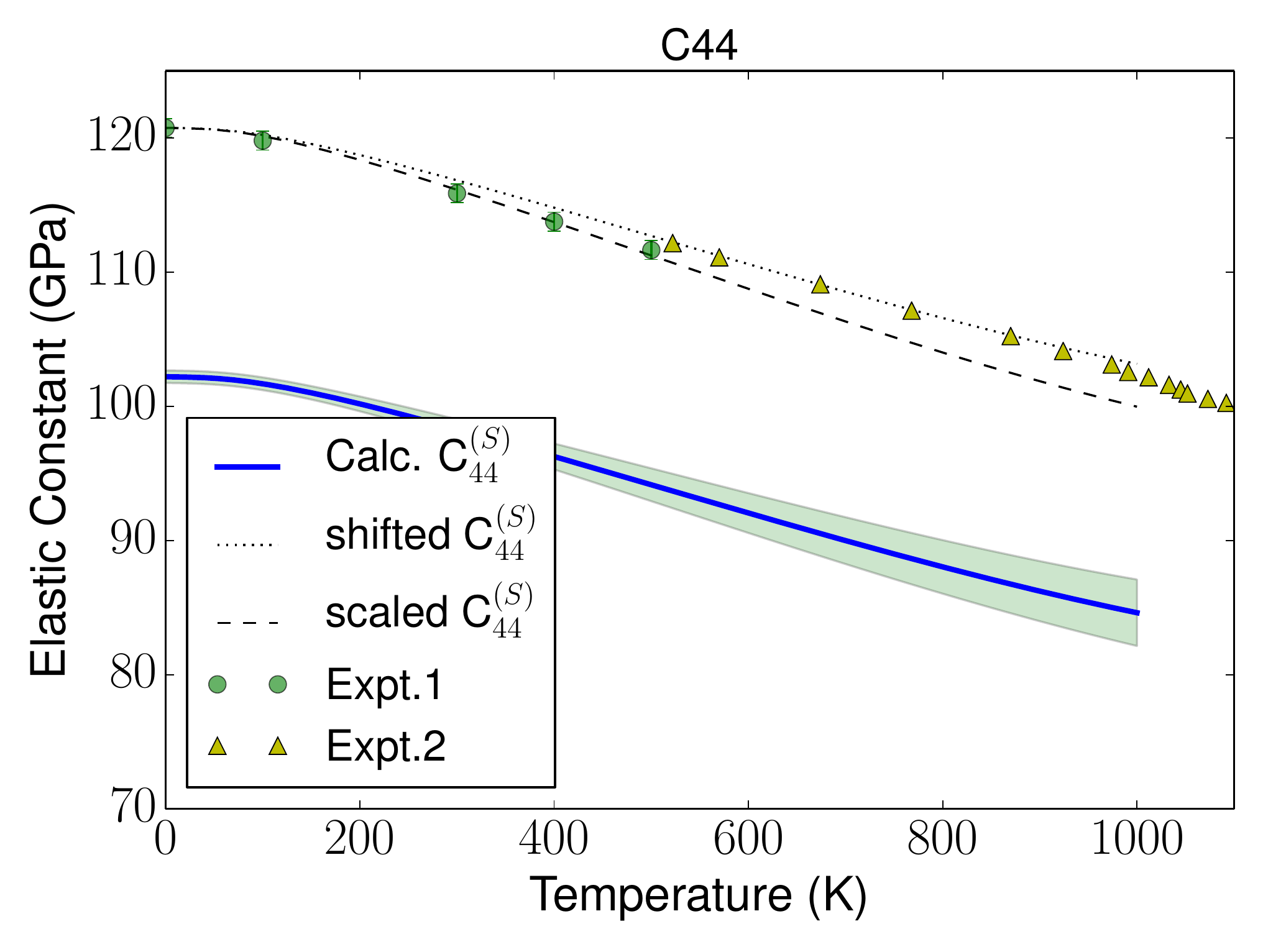} 
  \caption{Top panel: calculated adiabatic $C_{11}$ elastic constant (blue solid line). Bottom panel: calculated adiabatic $C_{44}$ (blue solid line). Two sets of experimental data are reported on each plot -- Expt.1 (green circles) from Ref.~\onlinecite{Adams} and Expt.2 (yellow triangles) from Ref.~\onlinecite{Dever}. The calculated interval of confidence is displayed as a shaded area. As a guide to the eye, we also plot the elastic constants rigidly shifted  (dotted line) and scaled (dashed line) to match the experimental values at zero temperature.}
  \label{fig:C11_C44}
\end{figure}

\begin{figure}
  \centering
  \includegraphics[trim=0mm 0mm 0mm 0mm, clip, width=0.52\textwidth]{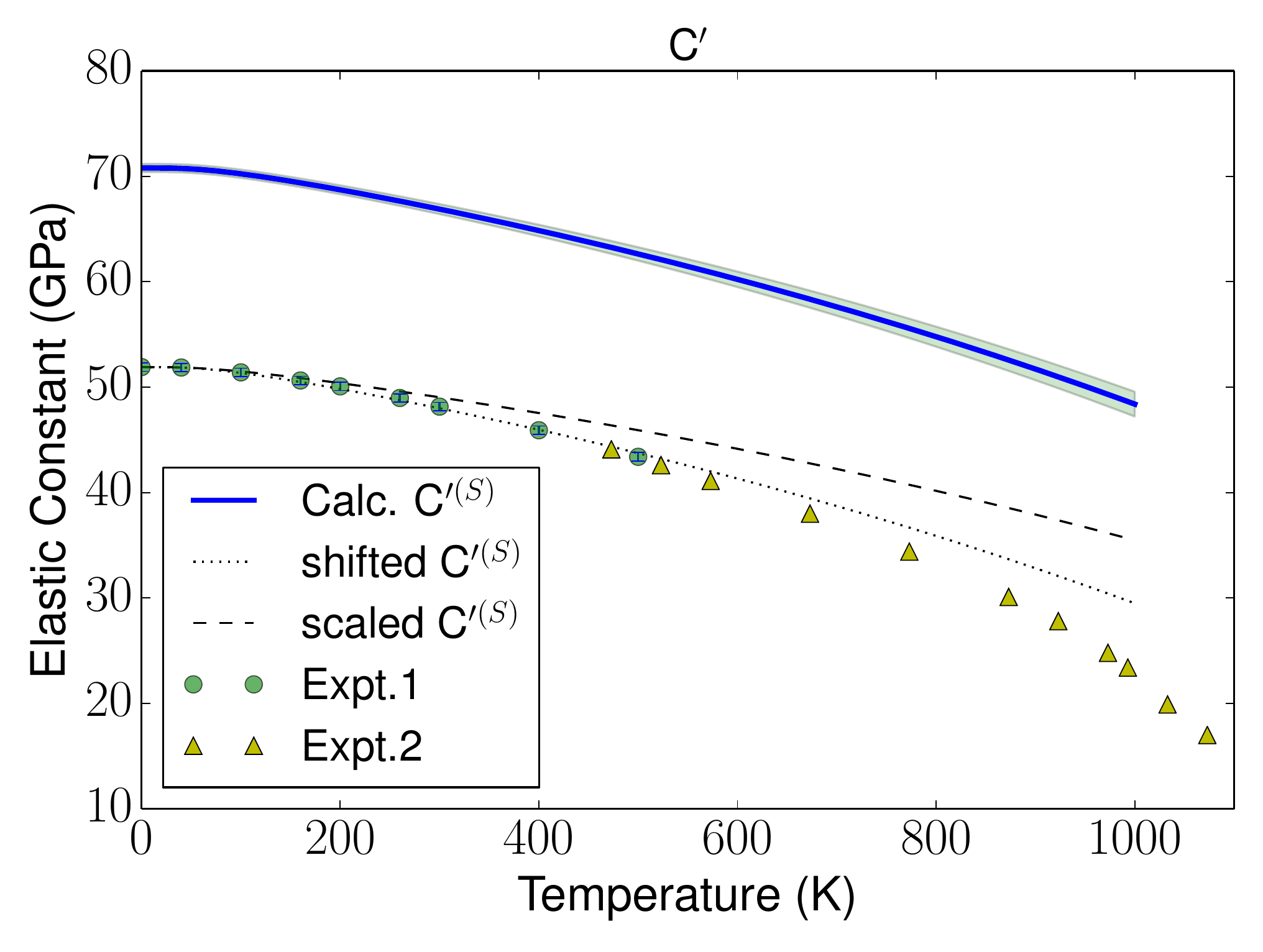} 
  \caption{Thermal behavior of the $C^{\prime}$ elastic constant calculated as a linear combination of $B(T)$ and $C_{11}(T)$ (blue continuous line). Two sets of experimental data are reported -- Expt.1 (green circles) from Ref.~\onlinecite{Adams} and Expt.2 (yellow triangles) from Ref.~\onlinecite{Dever}.  The calculated interval of confidence is displayed as a shaded area. As a guide to the eye, we also plot the elastic constants rigidly shifted  (dotted line) and scaled (dashed line) to match the experimental values at zero temperature.}
  \label{fig:Cprime}
\end{figure}

The calculated $C_{11}$ and $C_{44}$ elastic constants of BCC $\alpha$-iron 
both decrease by increasing temperature, as shown in Fig.~\ref{fig:C11_C44}. Our results are in reasonable  accordance with those reported in Ref.~\onlinecite{sha_cohen} (the exception is $C_{44}$ that in our case is fairly underestimated) where, however, a direct detailed comparison with experimental thermal softening is clearly more difficult.
In Tab.~\ref{tab:C0Kab}, we report their zero temperature values with and without zero-point energy (ZPE) contributions,  thus comparing these to experimental values. Also, for sake of completeness, we report in Tab.~\ref{tab:C0Kab_others} 
the $C_{12}=\frac{3B-C_{11}}{2}$, $C^{\prime}=\frac{1}{2}(C_{11}-C_{12})= \frac{3}{4}(C_{11}-B)$ and anisotropy ratio $C_{44}/C'$ obtained from 
standard theory of elasticity and our calculated $B$, $C_{11}$ and $C_{44}$ (see Fig.~\ref{fig:Cprime} for the temperature dependence of the $C^{\prime}$).
The inclusion of ZPE results in a small decrease of the elastic
constants and bulk modulus. The confidence interval at zero temperature is of the order of 0.1~GPa and
cannot account for the difference with respect to experiments, so we will discuss other possible
source of this discrepancy in the next section. 

\begin{table}[h]
\centering
\begin{tabular}{lllll}
\hline\hline
$T$ (K) &$a$ (\AA)& $B$ (GPa) & $C_{11}$ (GPa)  & $C_{44}$ (GPa)  \\ 
\hline
0 (no ZPE)& 2.834 &199.8$\pm$0.1 & 296.7$\pm$0.3 & 104.7$\pm$0.1 \\    
0 (ZPE)   & 2.839 &194.6$\pm$0.3 & 287.9$\pm$0.4 & 102.2$\pm$0.5 \\    
0 (Expt.)~\cite{Basinski,Adams} & 2.856 &170.3$\pm$1 & 239.5$\pm$1 &  120.7$\pm$0.1   \\
\hline
\end{tabular}
\caption{Calculated 0~K elastic constants for iron with and without zero-point energy contributions. Results are
compared to experimental data extrapolated to 0~K.}
\label{tab:C0Kab}
\end{table}

\begin{table}[h]
\centering
\begin{tabular}{llll}
\hline\hline
$T$ (K) & $C_{12}$ (GPa)& $C'$ (GPa)  & $C_{44}/C'$  \\ 
\hline
0 (no ZPE)&  151.4$\pm$0.2 &  72.7$\pm$0.3 & 1.44\\    
0 (ZPE)   & 148.01$\pm$0.5 &  70.0$\pm$0.4 &  1.46 \\    
0 (Expt.)~\cite{Adams}  & 135.7 &  51.9   & 2.32\\
\hline
\end{tabular}
\caption{$C_{12}$ and $C$' elastic constants, and $C_{44}/C'$ anisotropy ratio, derived from Tab.~\ref{tab:C0Kab} with and without zero-point energy.
Results are compared to experimental data extrapolated at 0~K. Errors are obtained according to propagation
of uncertainties.}
\label{tab:C0Kab_others}
\end{table}


\subsection{Discussion}
\label{sec:discussion}
The temperature dependence of the bulk modulus and of the elastic constants display an
overall good agreement with the available experimental data, showing how lattice
vibrations alone provide a robust description of the thermoelastic properties of the material, especially
below the Debye temperature $\Theta_D$. The agreement is still valid above $\Theta_D$ for the $C_{44}$,
that shows a near-linear behavior generally expected for metallic systems according to the
semiempirical Varshni equation~\cite{Varshni}. On the other hand, approaching the Curie temperature (1043~K) from below, the results are not able to reproduce the anomalous non-linear softening which is observed in experimental $B$ and $C_{11}$.

According to previous work, e.g. based on a tight-binding approximation
coupled to a single-site spin-fluctuation theory of band magnetism~\cite{Hasegawa}, effective spin-lattice
couples models~\cite{Rusanu} as well as experiments~\cite{Fultz,Dever}, the origin of these anomalies is inherently related to magnetic fluctuations and, ultimately, to their influence on the free energy landscape 
(via modulation of the exchange couplings, configurational disorder and magnon-phonon interaction).
In support of this conclusion, previous ab-initio papers~\cite{neugebauer,sha} suggest that the electronic 
entropy and phonon-phonon anharmonic effects beyond the quasiharmonic approximation play a minor role in determining the thermodynamics of the system below $T_C$. Given the strong indications of the 
pivotal role of magnetism in the description of thermoelastic properties of $\alpha$-iron close to the Curie temperature, further ab-initio calculations taking into account magnetic disorder would be of 
paramount interest (see for instance Ref.~\onlinecite{Neugebauer-phonon}).

Focusing instead on the 0~K structural and elastic properties, we now discuss the possible origin
of the mismatch between the calculated and experimental values. As we showed earlier, our calculated
 points are numerically accurate, and the errors associated with the fit are fairly small. As a 
consequence, the origin of the discrepancy can be ascribed (i) to the
presence of magnetic domain walls, (ii) to the pseudopotential approximation, (iii) to the approximate
XC functional.

First, we inspected the possible effects on the equilibrium volume at 0~K due to the presence of
magnetic domain walls. We focus our attention to the collinear domain wall case, thus mimicking a magnetic
distribution of domains in iron as two 8-atoms-thick ferromagnetic strips with antiparallel magnetic moments
repeated in the z direction through periodic boundary conditions. The effect due the interfaces is to increase the
lattice parameter of about 0.7$\%$ and to decrease the bulk modulus of about $9\%$. Since the density of
domain walls in real materials is expected to be an order of magnitude lower, the effect on the lattice
parameter should be rescaled accordingly, thus suggesting that domain walls affects only marginally
the value of the equilibrium lattice parameter at 0~K. 

Next, we have observed that, for a given XC functional, details of the pseudopotential can have a
large impact on the calculated quantities. For instance, in the case of PBE, 
the bulk modulus at 0~K ranges from 165 to 201 GPa if we consider pseudopotentials generated by different authors: ultrasoft or PAW, with 8 or 16 electrons or, even using the same electronic configuration but different version of \textsl{pslibrary}~\footnote{\protect\url{http://www.qe-forge.org/gf/project/pslibrary}} (see Figs.~\ref{graph:volume_hist},~\ref{graph:bulk_hist}).  

As discussed in Sec.~\ref{sec:compdetails}, the pseudopotential used in this work was chosen 
as being closest in its equation of state and its magnetization as a function of volume to all-electron FLAPW calculations~\cite{Cottenier}$\,$~\footnote{P. Pavone and S. Cottenier, private communications as reported above.}.
As a result, the discrepancy at 0~K with respect to experiments found in this work and in all-electron
calculations seems ascribable mainly to the exchange-correlation functional used.  
For this reason, we explored the effect of the XC functional on the 0~K properties, keeping the pseudopotential generation scheme and parameters unchanged. We performed test calculations with the WC~\cite{WC} and PBEsol~\cite{PBEsol} functionals, and found that the disagreement with the experimental data is increased (see Figs.~\ref{graph:volume_hist},~\ref{graph:bulk_hist}).

Eventually, we observe that the QHA thermal contribution to the energy landscape is almost linear (see Fig.~\ref{fig:Thermal}) and does not contribute too much to the total curvature in the energy landscape. Moreover, its change in second derivative along with temperature is even smaller, and only marginally contributes to the temperature dependence of the total curvature of the energy landscape (its main effect is to shift the minimum of the free energy as a function of temperature). Therefore, we conclude first that the mismatch with experiments at finite temperature is dominated by the 0~K static contribution discussed above. Second, that the temperature dependence of the elastic constants is driven, in first approximation, by the curvature of the 0~K energy landscape at the equilibrium expanded volumes.
Our finding suggests that one could try and employ more computationally expensive methods (such as DFT+U+J~\cite{CococcioniU, CococcioniUJ}, hybrid functionals~\cite{hybrids}, RPA~\cite{RPAdeGironc,RPA-Kresse} or DMFT~\cite{DMFT,Abrikosov-epsilon}) to explore possible improvements in the description of the 0~K mechanical properties of $\alpha$-iron, while thermal properties can be determined using lattice dynamics calculations performed with standard semi-local GGA functionals.

\section{Conclusions}
\label{sec:conclusions}
We have calculated the isothermal and adiabatic elastic constants of $\alpha$-iron as a function of
temperature from first-principles, using pseudopotential total energy calculations based on DFT and lattice dynamics
calculations based on DFPT, out of which we calculate free energies in the quasiharmonic approximation and finite-temperature
elastic constants from small strain deformations. Great care has been put in the
verification of the pseudopotentials, and the validation of the results
against experiments. 
Common semi-local DFT functionals such as PBE reproduce only fairly elastic constants at 
zero temperature; on the other hand, their thermal behavior, originating from the changes
in phonon dispersions upon crystal expansion, is very well described by the same functionals
and in the quasiharmonic approximation, with a softening of the elastic constants 
and bulk modulus that is in excellent agreement with experiments up to $\Theta_D$ and above. This work was supported by a grant from the Swiss National Supercomputing Centre (CSCS) under project ID \textit{ch3}.
We would like to acknowledge P. Pavone and C. Draxl for their kind support in calculating and providing all-electron data with the \protect{\exciting} code as well as S. Cottenier for providing all-electron data obtained with the \textsc{WIEN2k} code that were used for comparison with available pseudopotentials. We also acknowledge partial support from the FP7-MINTWELD project. 

\bibliography{APS_version}
\end{document}